%% file: main.tex
\documentclass[runningheads]{lmcs}
\pdfoutput=1

\usepackage{lastpage}
\lmcsdoi{22}{1}{8}
\lmcsheading{}{\pageref{LastPage}}{}{}%
{Jan.~23,~2025}{Mar.~05,~2026}{}

\usepackage[frozencache=true,cachedir=.//,newfloat,langlinenos=true]{minted}
\usepackage{float}

\usepackage{algorithm}
\usepackage{algpseudocode}

\usepackage{graphicx} 
\usepackage{array}    
\usepackage{booktabs} 
\usepackage{multirow} 
\usepackage{makecell} 

\usepackage{caption} 
\usepackage[utf8]{inputenc}

\input{macros}

\begin{document}

\title[Formal Analysis of {\tt CARE} with {\sc Uppaal}]{Formal Analysis of the Contract Automata Runtime Environment with {\sc Uppaal}: Modelling, Verification and Testing}



\author[D.~Basile]{Davide Basile\lmcsorcid{0000-0002-7196-6609}}[]

\address{Formal Methods and Tools Lab, ISTI--CNR, Pisa, Italy}	
\email{davide.basile@isti.cnr.it}  

\begin{abstract}
Recently, a distributed middleware application called contract automata runtime environment ({\tt CARE}) has been introduced to realise service applications specified using a dialect of finite-state automata.
In this paper, we detail the formal modelling, verification and testing of {\tt CARE}.
We provide a formalisation as a network of stochastic timed automata.
The model is verified against the desired properties with the tool {\sc Uppaal},
utilising exhaustive and statistical model checking techniques.
Abstract tests are generated from the {\sc Uppaal} models that are concretised for testing {\tt CARE}.
This research emphasises the advantages of employing formal modelling, verification and testing processes to enhance the dependability of an open-source distributed application.
We discuss the methodology used for modelling the application and generating concrete tests from the abstract model, addressing the issues that have been identified and fixed.
\end{abstract}

\maketitle

\section{Introduction}\label{sec:introduction}

Behavioural contracts~\cite{BCZ15} have been introduced to formally describe the interactions among 
 services, to enable to reason  formally about well-behaving properties of their composition. 
 Examples of properties are agreement among the parties or  reachability of 
 target states.

Contract automata are a dialect of finite state automata used to specify behavioural contracts formally  
in terms of offers and requests~\cite{BDF16}. 
A composition of contracts is in  {\it agreement} when all requests are 
matched by corresponding offers of other contracts.
A composition can be refined to one in agreement using 
 the orchestration synthesis algorithm~\cite{BBDLFGD20,BBP20}, 
 a variation of the synthesis algorithm from 
 supervisory control theory~\cite{RW87}.
%
%
%
%
The Contract Automata Runtime Environment ({\tt CARE})~\cite{DBLP:conf/fm/BasileB23} provides a middleware to coordinate the services implementing contracts. 
In {\tt CARE}, each transition of the orchestration automaton is executed by a series of interactions among the orchestrator and the {\tt CARE} services, implemented using Java TCP/IP sockets. 
These interactions may vary according to the specific configuration chosen among those provided by {\tt CARE}. 
In~\cite{DBLP:conf/fm/BasileB23}  the algorithms implemented in {\tt CARE} are proved to enforce  the adherence of each contract specification to its {\tt CARE} implementation (basically, the control flow of the application follows the synthesised orchestration automaton). 


 In this paper, we describe the modelling, verification and testing of the low-level interactions  among {\tt CARE} services and the orchestrator. 
This aspect  is no less important, as witnessed by known cases  of algorithms proved to be correct (e.g., the Byzantine distributed consensus~\cite{DBLP:journals/toplas/LamportSP82}) and whose low-level communication implementations were found to have issues, e.g. deadlocks~\cite{Roggenbach2022}.
We verify several properties, including  the absence of deadlocks, absence of undelivered messages, and reachability of target states. 

The formal model is a network of stochastic timed automata as accepted by the {\sc Uppaal} toolbox. 
Different variants of the formal model are proposed.  
The model undergoes verification against the desired properties using {\sc Uppaal}, employing both exhaustive model checking, statistical model checking and model-based testing. 
This combination allows for thorough analysis of the model's behaviour and ensures scalability when dealing with systems that possess a large state-space.

The benefits of modelling and verifying {\tt CARE} are: 
(i)~increasing the  confidence in the reliability of {\tt CARE} due to the formal verification through model checking and model-based testing, 
(ii)~quantitative evaluation of measures of interest of {\tt CARE} obtained through statistical model checking,
(iii)~improved documentation thanks to the graphical and animatable state machines endowed with precise semantics. 

Finally, this paper tackles the challenge of providing a full-fledged model-based development and formal methods approach~\cite{GBP20}. The final application has been graphically modelled at an abstract level, formally verified and tested using the formal model. 
The criteria followed for abstracting away irrelevant details are discussed together with the issues that have been found and fixed thanks to the formal modelling, verification and testing. 

All models, formulas, and logs are publicly available in~\cite{UppaalModels}, together with traceability and model-based testing information connecting the model to the source code.

This work is an extension of the conference paper~\cite{DBLP:conf/coordination/Basile24}. 
The extension includes the addition of more figures and tables, as well as further explanations throughout the paper. The most significant addition concerns the testing phase, which was only briefly mentioned in~\cite{DBLP:conf/coordination/Basile24}. Testing is now thoroughly described in Section~\ref{sect:testing}. 
We cover both the generation of abstract tests from the {\sc Uppaal} model and the process of concretising these tests in the real application.

Summarising, the core contributions of this paper are:
    (i) the bottom-up formal modelling of {\tt CARE},  an already established, open-source distributed middleware, modelled as a network of stochastic timed automata suitable for the {\sc Uppaal} toolbox;
    (ii) the verification of desired properties of the {\tt CARE} formal model through the application of both exhaustive and statistical model checking techniques within {\sc Uppaal};
    (iii) the establishment of a direct connection between the abstract formal model and the actual source code of {\tt CARE} using traceability and model-based testing. Abstract tests are generated from the {\sc Uppaal} models and concretised as JUnit tests for the {\tt CARE} implementation. This connection helps to validate the chosen level of abstraction. 

{\bf Outline} 
We start by discussing the related work in Section~\ref{sect:relatedwork}. 
In Section~\ref{sect:background}, we provide the background on contract automata, their tool support, {\sc Uppaal}, and the contract automata runtime environment. 
The methodology used for modelling and analysing {\tt CARE} is described in Section~\ref{sect:methodology}. 
Section~\ref{sect:formalmodel} contains the description of the models, whilst the verification is described in Section~\ref{sect:analysis}. 
Model-based testing is addressed in Section~\ref{sect:testing}. 
Finally, the conclusion and future work are presented in Section~\ref{sect:conclusion}.

\section{Related work}\label{sect:relatedwork}

Several applications of {\sc Uppaal} to various case studies are available in the literature, including land transport~\cite{BBL20}, maritime transport~\cite{DBLP:conf/sefm/Shokri-Manninen20}, medical systems~\cite{DBLP:journals/corr/abs-2203-09884}, and autonomous agents path planning~\cite{DBLP:journals/sttt/GuJPSEL22}. 
These case studies, along with the present paper, adopt a model-based approach, wherein partial  representations of the applications are created using models.

A recent survey conducted on formal methods~\cite{10.1145/3520480} reveals that numerous publications lack a direct correlation between the concrete implementation and its abstract model.  
This holds especially when the systems are only envisioned, yet to be realised, such as for example the ERTMS~L3 railway signalling system~\cite{BBC18,DBLP:journals/sttt/BasileBFL22}. 
This is also the case for industrial systems whose implementations are non-disclosed~\cite{sys.21679}. 
In all these cases, determining the accuracy of the model in relation to the actual system and assessing the appropriateness of the chosen level of abstraction becomes challenging.  
Furthermore, measuring the influence of the formal modelling and verification phase on the analyzed system poses difficulties. 

Indeed, a recent survey on formal methods in the transport domain~\cite{10.1145/3520480} identifies the need 
``to lower the
degree of abstraction, by applying formal methods and tools to development phases that are closer to software
development''. 
This is also confirmed by a recent survey on formal methods among 130 high-profile experts~\cite{GBP20}, where nearly half of them believe that the most likely future users of formal methods are ``a small number of skilled experts'' as long as ``skills like modelling, specification,  abstraction are involved''.

In contrast to the previously mentioned literature, in this paper we present a bottom-up formal analysis of an established open-source system that has already been developed~\cite{DBLP:conf/fm/BasileB23}. 
The availability of the source code further enables us to establish a connection between the abstract  formal model and the actual source code. This capability facilitates the precise identification of specific aspects of the real system that have been abstracted in the formal model. Additionally, it allows us to validate the appropriateness of the chosen level of abstraction. 

Recently, Yggdrasil~\cite{DBLP:conf/fortest/HesselLMNPS08}, the offline model-based testing functionality of {\sc Uppaal}, has been used in~\cite{DBLP:conf/fmics/KimLNMO15} for testing an industrial real-time automotive turn indicator system, and in~\cite{DBLP:conf/medi/HammamiL23} for testing EIP-1559 Ethereum smart contracts.
The usage of the {\sc Uppaal} suite, and more specifically Yggdrasil, in three industrial case studies (i.e., medical devices, an automotive system and a pump manufacturing) is reviewed in~\cite{DBLP:conf/isola/LarsenLN18}.
Many of the above applications of model-based testing using {\sc Uppaal} concern industrial, non-disclosed case studies.
Therefore, only the models and abstract test cases are reported, whilst concrete test cases and source code have not been made public.
In comparison, {\tt CARE} is an open-source application. The concrete tests generated from the {\sc Uppaal} models discussed in this paper, as well as the system under testing, are available at~\cite{UppaalModels}.
To the best of our knowledge,  there are no other non-trivial open-source applications for which their {\sc Uppaal} formal model is openly accessible and directly connected to the source code through traceability and model-based testing. 
This contribution assists in linking formal methods, particularly {\sc Uppaal}, to the software development process.

The contract automata approach is closer to~\cite{DBLP:journals/scp/KouzapasDPG18}, where behavioural types are expressed as finite state automata of {\tt Mungo}, called typestates~\cite{DBLP:journals/tse/StromY86,DBLP:conf/coordination/AlsubhiD24}.
Similarly to {\tt CARE}, in {\tt Mungo} finite state automata are used as behaviour assigned to Java classes (one automaton per class), with transition labels corresponding to methods of the classes.
A tool to translate typestates into automata was presented in~\cite{DBLP:journals/corr/abs-2009-08769}. {\tt CATApp} is a graphical front-end tool for designing contract automata~\cite{CATAPPurl}.
A tool similar to {\tt Mungo} is {\tt JaTyC} (Java Typestate Checker)~\cite{BACCHIANI2022102844}.

Other model-based verification approaches for service-oriented systems and distributed middleware are surveyed in~\cite{Rai_Gangadharan_2021,10.1145/3595376}. 
In the literature, there exist many formalisms for modelling and analysing the behaviour of services, ranging from behavioural type systems, including behavioural contracts~\cite{CGP09,ABZ13,LP15} and session types~\cite{BLMT08,HYC08,DD09,CDP12,MNF13}, to automata-based formalisms, including interface automata~\cite{AH01} and (timed) (I/O) automata~\cite{LT89,AD94,DLLNW10}. Foundational models for service contracts and session types are surveyed in~\cite{BBG07,BCZ15,Hut16}.
Sessions and session types~\cite{BLMT08,HYC08,DD09,CDP12,MNF13} have been introduced to reason over the behaviour of services in terms of their interactions. 
Compared to contract automata, behavioural contracts and session types use process-algebraic frameworks, instead of finite state automata. Indeed, contract automata are similar to other software engineering-based  formalisms such as I/O automata~\cite{LT89,AD94,DLLNW10}. 

In the literature there are various attempts to model applications that, similarly to {\tt CARE}, communicates through TCP/IP sockets, also using {\sc Uppaal}~\cite{DBLP:journals/ijcomsys/WangWHLGCG22,DBLP:conf/tase/FeiZL18,DBLP:journals/corr/SainiF17,WITSCH2006101}. In~\cite{DBLP:conf/secdev/CluzelGMZ21} the verification of the transport layer of a TCP/IP protocol stack implementation in C is performed with a  translation into SPARK, where contracts are modelling the corresponding behaviour. 
A method to obtain SPIN formal models from TCP/IP socket applications implemented in C is presented in~\cite{DBLP:journals/sttt/CamaraGMS09} where the C application is embedded into a PROMELA model  equipped with a formalisation of the underlying TCP/IP socket API. Similarly, we provide a model of the underlying socket communications but we also modelled aspects of the application that are targeted by our analysis whilst unnecessary details are abstracted away. 
Another approach~\cite{DBLP:conf/nfm/LiuSMO020} presents an automatic translation from MAUDE specifications of distributed applications to  distributed MAUDE implementations using TCP/IP sockets, where the implementations are formally proven to be in correspondence with the specifications. 
Our approach is bottom-up and is based on {\sc Uppaal} modelling of an already realised application. 

As reported in~\cite{Boulanger}, the question of qualification and validation of formal methods tools is ``absolutely crucial''. 
Other tools for behavioural contracts  are present  in the literature~\cite{Gay20171,DBLP:conf/facs2/OrlandoPBLT21,DBLP:conf/coordination/PomboMT24}, but differently from {\tt CARE}, many of these implementations have not undergone a  process of formal verification.

\section{Background}\label{sect:background}

In this section, we will provide background on contract automata, their software support, and the {\sc Uppaal} statistical model checker.  
The focus of this paper is on the formal analysis of the runtime environment of contract automata.
While we offer a concise overview of contract automata to enhance comprehension of their runtime environment, they are not the  focus of our formal analysis.

\subsection{Contract Automata}
\begin{figure}[t]
    \centering
\includegraphics[width=0.49\textwidth]{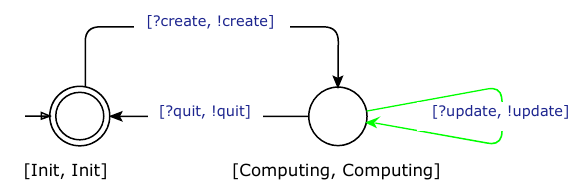}
    \vspace*{-\baselineskip}\caption{An example of contract automaton}
    \label{fig:contractautomaton}
\end{figure}

Contract automata are a dialect of finite-state automata modelling services that exchange offers and requests.
A contract automaton models either a single service or a composition of interacting services. 
Labels of transitions of contract automata are vectors of atomic elements called \emph{actions}.  Similarly, states of contract automata are vectors of atomic elements called basic states.
The length of the vectors is equal to the number of services in the automaton (this number is called \emph{rank}). 
A request (resp., offer) action is prefixed by~{\tt ?} (resp.,~{\tt !}). 
The idle action is denoted by~{\tt -}.
Labels are constrained to be one out of three types. 
In a \emph{request}  (resp., \emph{offer}) label a service performs a request (resp., offer) action  and all other services are idle. 
In a \emph{match} label one service performs a request action, another service performs a matching offer action, and all other services are idle.  
For example, the contract automaton in Figure~\ref{fig:contractautomaton} right has rank~2 and the label {\tt [?quit,!quit]} is a match where the request action {\tt ?quit} is matched by the offer action {\tt !quit}. 
Note the difference between a request label, e.g. {\tt [?coffee,~-]},  and a request action, e.g. {\tt ?coffee}.

In a composition of contracts various properties can be analysed~\cite{BDF16}.
For example, the property of \emph{agreement\/} requires to match all request actions, whereas offer actions can remain unmatched. 
The synthesis of the orchestration in agreement produces a sub-automaton of the composition where all services can match their requests with corresponding offers to reach a final state. 
Thus, in the orchestration in agreement labels of transitions are only matches or offers~\cite{BBP20}. 
The contract automaton in Figure~\ref{fig:sockettimeoutandorchestration} right is an orchestration in agreement. 
%
%

Contract automata and their functionalities are implemented in a software artefact, called Contract Automata Library (\texttt{CATLib})~\cite{BB22OSP}.
With {\tt CATLib} it is possible to specify, compose, and synthesise specifications given as contract automata. 
Indeed, {\tt CARE} uses the facilities offered by {\tt CATLib} for, among other things, composing contracts and synthesising orchestrations.
This software artefact is a by-product of our scientific research on behavioural contracts and implements results that have previously been formally specified in several publications (cf., e.g.,~\cite{BBDLFGD20,BBP20,BDF16}). 
 Scalability features offered by {\tt CATLib} include a bounded on-the-fly state-space generation optimised with pruning of redundant transitions and parallel streams computations~\cite{DBLP:journals/corr/abs-2308-10651,DBLP:journals/jlap/BasileB24}. 
 The software is open source~\cite{BB22OSP}, it has been developed using principles of model-based software engineering~\cite{BasileB21} and it has been extensively validated using various testing and analysis tools to increase the confidence on the reliability of the library~\cite{BB22OSP}.

\subsection{Uppaal}
\textsc{Uppaal} SMC~\cite{DLLMP15} is a variant of \textsc{Uppaal}~\cite{BDLHPYH06}, which is a well-known toolbox for the verification of real-time systems. 
%
\textsc{Uppaal} models are stochastic timed automata, in which non-determinism is replaced with probabilistic choices,  and time delays with probability distributions (uniform for bounded time and exponential for unbounded time). 
These automata may communicate only via broadcast channels and shared variables. 
In this paper, we will use both exhaustive and statistical model checking. 
Statistical Model Checking (SMC)~\cite{LLTYSG19} involves running a sufficient number of (probabilistic) simulations of a system model to obtain statistical evidence (with a predefined level of statistical confidence) of the quantitative properties to be checked. 
Monte Carlo estimation with Chernoff-Hoeffding bound executes   
$N = \lceil (ln(2)-ln(\alpha))/(2\epsilon^2) \rceil$ 
simulations $\rho_i$, $i\!\in\!1...N$,  
to provide the interval $[p'-\epsilon,p'+\epsilon]$ 
 with  confidence  $1-\alpha$.
Here,  $p'=(\#\{\rho_i\!\mid\!\rho_i\!\models\!\varphi \}) / N $,
 i.e., $\mathtt{Pr}(|p'-p|\leq \epsilon)\geq 1 - \alpha$ where $p$ is the unknown value of the formula $\varphi$ being estimated statistically and $\epsilon$ and $\alpha$ are the user-defined precision and confidence, respectively. 
Crucially, the number of simulations used to estimate a formula is independent of the model's size and depends only on the parameters $\alpha$ and $\epsilon$. 
 In practice the number of simulations required by {\sc Uppaal} to reach a specific confidence level is optimized and is thus lower than the above theoretical bound.  
  {\sc Uppaal} supports template automata used to instantiate different copies (in different experiments) of the same automaton, 
distinguishable by their parameters.  
 The selection of {\sc Uppaal} as the chosen tool is influenced by several factors. These include its extensive adoption by the community, expertise of the author, usability, primitive support for real-time and stochastic modelling, probabilistic and non-deterministic choices, and capabilities for statistical model checking, simulation, and model-based testing.
 
The off-line test case generator of {\sc Uppaal} is called Yggdrasil, which is integrated since version 4.1 into the {\sc Uppaal} GUI as a tab.
Starting from the {\sc Uppaal} models, a suite of test cases is created, aiming to cover all syntactic transitions in the model (i.e., edge coverage). The {\sc Uppaal} model can be decorated with test code, which can be generated from the execution of a transition or at the entry to or exit from a location. Any language can be used for the test code. 
An abstract test case is generated from a trace of execution of the {\sc Uppaal} model. During the execution of the trace, the test code of the model is dumped into a text file, forming the abstract test case. 
When the system is composed of a set of automata, it is possible to generate the test code from a single automaton or from the whole system. In this latter case, the single generated abstract test case contains the interleaved test code from all automata.
The traces of execution can be generated as witnesses of some reachability property defined using the {\sc Uppaal} syntax.
Random traces can also be executed to improve test coverage.
Parameters to be set include the depth of the trace, the type of search (i.e., breadth, depth), and options for the trace to be generated (i.e., fastest, shortest).
The generated abstract test cases need to be decorated and filled with the missing information from the system under test to become concrete test cases. 
More details will be provided in Section~\ref{sect:testing}. 

\subsection{CARE}
\begin{figure}[t]
    \centering
    \includegraphics[width=.95\columnwidth]{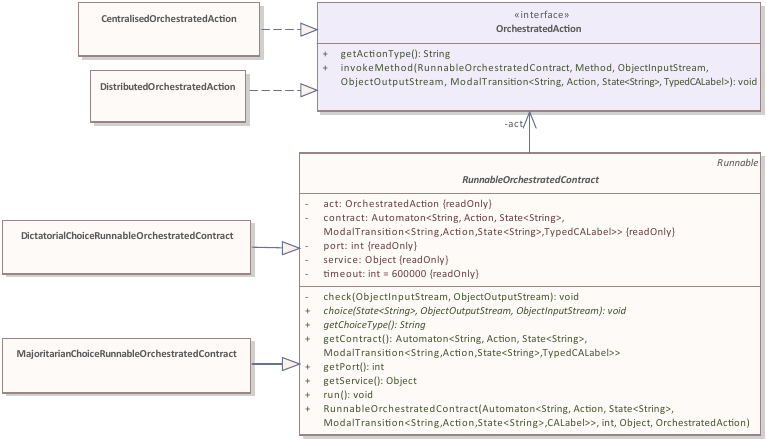}
    \caption{The class diagram for the orchestrated services;
the methods of the derived classes are visible in their super-class/interface as abstract methods (in italic)}
    \label{fig:orchestrated}
\end{figure}

\begin{figure}[t]
    \centering
    \includegraphics[width=.95\columnwidth]{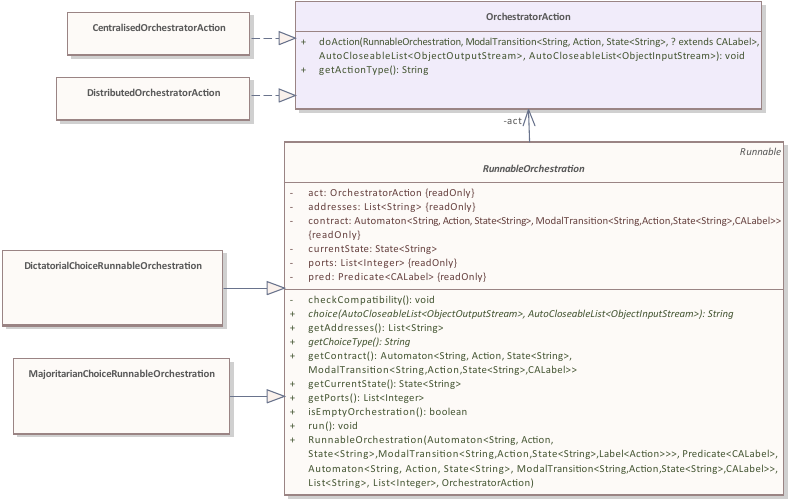}
    \caption{The class diagram for the orchestrator;
the methods of the derived classes are visible in their super-class/interface as abstract methods (in italic)}
    \label{fig:orchestration}
\end{figure}

The Contract Automata Runtime Environment ({\tt CARE})~\cite{DBLP:conf/fm/BasileB23} provides facilities for pairing the contract automata specifications with actual implementations of service-based applications. 
Note that our purpose is not to formally analyse the applications created using {\tt CARE}, but to formally analyse {\tt CARE} itself. 
The formal verification of applications developed using {\tt CARE} is a consequence of the reliance of {\tt CARE} on the formal guarantees provided by contract automata~\cite{DBLP:conf/fm/BasileB23}, along with the formal verification of {\tt CARE} as an application.  This paper addresses the latter aspect.
This software is organised into classes for the orchestrated services (cf.\ Figure~\ref{fig:orchestrated}) and classes for the orchestrator (cf.\ Figure~\ref{fig:orchestration}). 
The two core abstract Java classes are the {\tt RunnableOrchestration} and the {\tt RunnableOrchestratedContract}.

\paragraph{Services}
In Figure~\ref{fig:orchestrated}, \texttt{RunnableOrchestratedContract} is an abstract class that implements an executable wrapper 
 responsible for pairing its contract automaton specification (instance variable {\tt contract} storing a contract automaton)   with an implementation provided as a Java class (instance variable {\tt service} implementing the service), where each action of the automaton {\tt contract} is in correspondence with a method of the {\tt service} class. 
The types of the methods of a matching offer and request must also be in correspondence.
Each time a new orchestration involving the service is initiated, the {\tt RunnableOrchestratedContract} creates a new service. This service remains in a waiting state to receive commands from the orchestrator, which then triggers the execution of the corresponding methods. 
In case the orchestrator requires to perform an action not prescribed by its service contract, then a {\tt ContractViolationException} will be raised by the service.  
As demonstrated in~\cite{DBLP:conf/fm/BasileB23}, this exception will never be raised if the orchestration in agreement is synthesised using the algorithms of contract automata.  



The realisation of an orchestration is abstracted away in contract automata. 
Crucially, offers and requests of contracts are an abstraction of low-level messages sent between the services and the orchestrator to realise them.
{\tt CARE}  exploits the abstractions provided by Java to allow its specialisation according to different implementation choices, using abstractions of object-oriented design, as showed in Figure~\ref{fig:orchestrated} and Figure~\ref{fig:orchestration}.

%


\paragraph{Orchestrator} The class diagram of the orchestrator side in  Figure~\ref{fig:orchestration}.
The abstract class \texttt{RunnableOrchestration} implements a special service that reads the synthesised orchestration (stored in the instance variable {\tt contract}) and orchestrates the services (the instances of \texttt{RunnableOrchestratedContract}) to realise the overall application.  The instance variables {\tt port} and {\tt addresses} contain the corresponding addresses and ports of the services involved in the orchestration.

\paragraph{Choices and Actions} Two  aspects to implement for both the orchestrator and the services are choices and termination (through the abstract method {\tt choice}).
Indeed, during the execution of an orchestration, the selection of the transition to execute in the presence of multiple enabled transitions is implemented through this abstract method.
{\tt CARE} is equipped with default implementations, but can be extended (by implementing the relative interfaces and abstract methods) to include other options, other than the default ones.
Currently, a so-called `dictatorial' choice (i.e., an internal choice of the orchestrator, external for the services) and a so-called `majoritarian' choice (services vote and the majority wins) are two implemented options. 
For the services, {\tt MajoritarianChoiceRunnableOrchestratedContract} and {\tt DictatorialChoiceRunnableOrchestratedContract} are the two classes specialising {\tt RunnableOrchestratedContract} according to how the choice is handled and implementing the abstract methods, whilst {\tt MajoritarianChoiceRunnableOrchestration} and {\tt DictatorialChoiceRunnableOrchestration} are specialising {\tt RunnableOrchestration}.

{\tt CARE} also provides default implementations for  the low-level message exchanges. 
Currently, the two available options are the `centralised' action, where the orchestrator acts as a proxy, 
and the `distributed' action, where two  services matching their actions directly interact with each other  once the orchestrator has made them aware of a matching partner and its address/port. 
Accordingly, each {\tt RunnableOrchestratedContract} has an {\tt OrchestratedAction}, and {\tt RunnableOrchestration} has an {\tt OrchestratorAction} (instance variable {\tt act}) used to implement the corresponding actions that can be either distributed or centralised according to the current implementation.


\begin{exa}
We discuss an example of interactions between two services and an orchestrator. 
The example comprehends both the {\tt CentralisedAction} and {\tt DistributedAction} implementations of {\tt CARE}. 
The sequence of service invocations is displayed in Figure~\ref{fig:actions}.
In this example, the orchestrator is executing a transition of the orchestration automaton that is labelled with  the match  \texttt{[?coffee,!coffee]}, in which \texttt{Alice} is requesting a \texttt{coffee} and \texttt{Bob} is offering a \texttt{coffee}.
As stated above, the  {\tt RunnableOrchestratedContract} class of both {\tt Alice} and {\tt Bob} has two instance variables ({\tt contract} and {\tt service}, resp.) containing the contract automaton object and the implementation provided as a Java class. 
In this example, the implementation class of {\tt Alice} has a method {\tt Integer coffee(String arg)} that is paired with the corresponding request action {\tt ?coffee}  of her contract. Similarly, the implementation class of {\tt Bob} has a method {\tt String coffee(Integer arg)} that is paired with the corresponding offer action {\tt !coffee}  of his contract. 
\begin{figure}[t]
    \centering
    \includegraphics[width=0.28\textwidth]{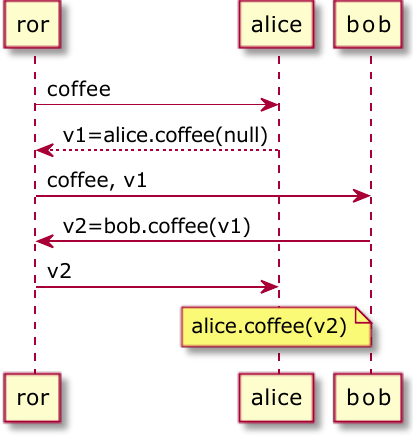} \qquad\qquad
    \includegraphics[width=0.38\textwidth]{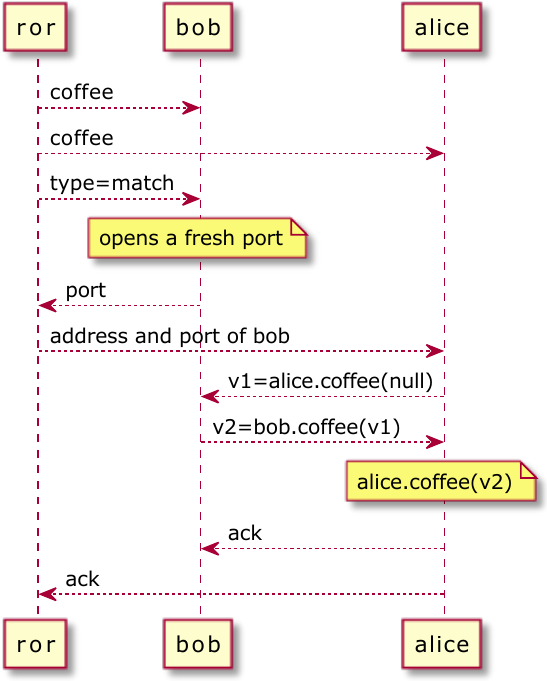}
    \caption{The sequence diagrams depicting the interactions implementing a centralised match action (on the left) and a distributed match action (on the right)}
    \label{fig:actions}
\end{figure}
%
%

In both centralised and dictatorial configurations, the method \texttt{coffee} of \texttt{Alice} is invoked twice: firstly, passing no argument, it returns an {\tt Integer} value (e.g., the amount of sugar) that is passed (by the orchestrator {\tt ror}) as an argument to the method \texttt{coffee} of \texttt{Bob}, which in turn returns a {\tt String} value that is eventually passed as an argument to the method \texttt{coffee} of \texttt{Alice}, thus fulfilling the {\tt coffee} request.

In the {\tt CentralisedAction} implementation, the orchestrator acts as a broker, forwarding to the involved parties the various invocations and responses.
In the {\tt DistributedAction} implementation, the orchestrator communicates the chosen action to both matching services, and the offerer is also notified of a match (rather than an offer) action, so that it can open a fresh port to interact with the requester, which is notified of the address and port of the offerer.
Upon successful termination of the interactions, each \texttt{RunnableOrchestratedContract} updates its contract status, and the orchestrator proceeds with the next invocation according to the overall orchestration contract, whose status is also updated.
In Section~\ref{sect:formalmodel} will discuss how this kind of interaction is modelled.


\end{exa}


\section{Methodology}\label{sect:methodology}

In this section, we describe both the methodology used for modelling the communications, abstracting away irrelevant details, and the adequacy of the model to the actual implementation. 

\subsection{Modelling TCP/IP sockets communication}\label{sect:modellingtcpipsockets}
\begin{figure}[t]
    \centering
    \includegraphics[width=\columnwidth]{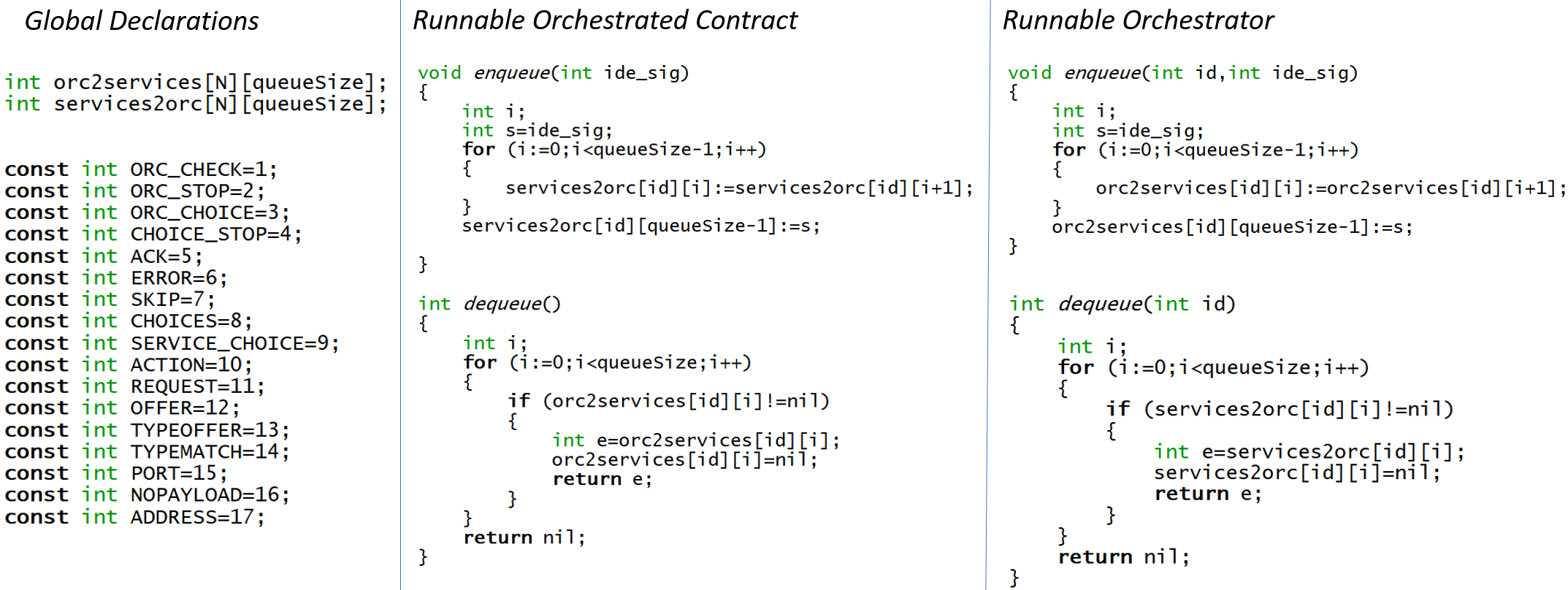}
    \caption{The {\sc Uppaal} code used to model the TCP/IP socket communications}
    \label{fig:tcpipmodel}
\end{figure}

 Java TCP/IP sockets communications are asynchronous with FIFO buffers~\cite{ObjectOutputStream}. 
 In {\sc Uppaal}, the interactions are via channel synchronisation and global variables.   
 Thus, TCP/IP sockets are solely employed in the actual implementation and are not utilized by the automata of the model. 
Instead, in the model each party communicates with the partner using two global arrays (one for sending and one for receiving, respectively). 
Figure~\ref{fig:tcpipmodel} shows the code used in the {\sc Uppaal} models (see Section~\ref{sect:formalmodel}) for modelling the TCP/IP socket communications. 

The two global (FIFO) arrays in the model are named {\tt orc2services[N][queueSize]} and {\tt services2orc[N][queueSize]} (see Figure~\ref{fig:tcpipmodel} left). These are used to model the TCP/IP socket buffers, where {\tt N} is the number of services in the model and {\tt queueSize} is the size of the queue of the components. 
These arrays are only modified by functions for enqueueing and dequeuing  messages.

The local declarations of the two automata contain methods for sending and receiving from the partners and checking their queues of messages. 
Both automata declare a method {\tt enqueue} for sending to the partner a message. 
Messages are modeled as global integer constants, depicted  in Figure~\ref{fig:tcpipmodel} left.  
Indeed, in this model the actual payload of each communication is abstracted away. 
For example, in Figure~3 left, the constants {\tt REQUEST} and {\tt OFFER} abstract payloads that could be any class instantiated by the user, comprehending newly created Java classes (see Section~\ref{sect:abstractions}).
The {\tt enqueue} method of the services only takes as parameter  the signal {\tt ide\_sig} to send  (see Figure~\ref{fig:tcpipmodel} center).  
The identifier of the service {\tt id} is only needed  on the orchestrator side to identify the partner (there is only one orchestrator thus no identifier is needed for it, see Figure~\ref{fig:tcpipmodel} right).
Similarly, both automata have a method {\tt dequeue} for consuming messages from their respective arrays. 

The default mode for Java TCP/IP sockets is \emph{blocking}~\cite{JavaSocket}, meaning that the sender blocks when the buffer of the receiver is full, and waits until there is enough space to proceed. 
Accordingly, a transition having a send in its effect will check in its guard whether there is enough space left in the array of the partner by calling  either the method {\tt available} (returning the space left) or {\tt isFull}.
Similarly, in Java TCP/IP sockets the read operation is \emph{blocking}. Accordingly, before reading it is always checked whether the array is not empty with the method {\tt !isEmpty}. When the array is empty the automaton blocks  until a message is received. 

Source locations of sending and receiving transitions are neither committed nor urgent. 
In fact, an enabled committed transition (i.e., whose source location is committed, denoted with {\tt C}) must be executed before any other non-committed transition in the network. 
Instead, an urgent transition must be executed without any delay.
If a sending or receiving transition were to be either committed or urgent, it could introduce the possibility of false positive deadlocks. This scenario arises when, for example, the receiver has a full buffer (i.e., array) and is prepared to free it, but the sender transition is enabled and committed.
Similarly, this false positive can occur when the buffer is full, the send transition is urgent  but the receive operation is not urgent, or vice versa, when the buffer is empty, the read transition is urgent, and the send operation is not urgent.

Hence, the operations of writing to and reading from a buffer are represented using stochastic delays, specifically following an exponential distribution. Two rates are employed to capture the delay associated with reading and writing. These exponential delays (variables {\tt write} and {\tt read}) are present in all non-committed and non-final locations of both automata in Figure~\ref{fig:runnableorchestration} and Figure~\ref{fig:runnableorchestratedcontract} (modelling the orchestrator and the services, cf. Section~\ref{sect:formalmodel}). 
However, the presence of unbounded delays introduces scenarios where the receiver (resp., sender) may wait indefinitely without executing its read (resp., write) transition, even when it is enabled. These scenarios  would invalidate the exhaustive model checking of reachability properties that are satisfied by the actual system, leading to false positives.
In the real implementation, Java TCP/IP sockets offer a timeout mechanism wherein an exception is thrown if no message exchange occurs within a specified time frame. All sockets used by {\tt CARE} have this timeout.
Consequently, a dedicated automaton called {\tt SocketTimeout} is replicated for each service to model the timeout operation (see Figure~\ref{fig:sockettimeoutandorchestration} left). Each send or receive operation in every socket resets the {\tt SocketTimeout} clock {\tt c}. If no reset operation is received within a certain duration (variable {\tt timeout}), {\tt SocketTimeout} enters a location  called {\tt Timeout} and broadcasts the signal {\tt fail}, indicating that an exception has been thrown. 
All automata have a transition from every location (except for {\tt Terminated}) to a location {\tt Timeout} that is reached upon receiving the signal {\tt fail}. For readability, these {\tt Timeout}  locations and all their incoming fail transitions are not shown in Figure~\ref{fig:runnableorchestration} and Figure~\ref{fig:runnableorchestratedcontract}.
 
\subsection{Abstractions}\label{sect:abstractions}
We have discussed the modelling of Java TCP/IP socket communications. 
We now discuss other aspects that have been abstracted away in the model. 
We note that the abstracted aspects are irrelevant for the analysis discussed in Section~\ref{sect:analysis}. 
We remark that the paper objective is to  verify the interactions of the {\tt CARE} middleware itself. 
Therefore, the underlying application that is executed by {\tt CARE} is abstracted away. 
Therefore, in the model, the underlying orchestration automaton is  abstracted together with the contracts of the services. 
Thus, all conditionals that are dependent from the underlying orchestration are abstracted as probabilistic choices. 
We assume that the orchestration has been correctly synthesised from the services contracts.  
This allows us to verify the interactions for any possible valid orchestration. 
If a specific orchestration would be modelled, then we would lose such generality. 
In particular, the payloads of the communications (e.g., which specific action, which choices) are abstracted. 
The conditions used to decide whether to perform a choice, an action or to stop are also abstracted away. 
The only  modelled condition is that no two consecutive choices are allowed (i.e., after choosing, the chosen step must be performed). 
When executing a transition, the conditions used to check whether the label of the transition is an offer or a match are abstracted away in the model.   
Moreover, the identifiers of the services involved in performing a choice or an action (which are concretely extracted from the labels of the transitions) are also chosen non-deterministically. Indeed, all services are distinguished replicas of the same automaton. 
Finally, we model a single orchestration. 
In fact, when multiple orchestrations are executed, they operate independently of each other and can be verified individually.

As stated above, the underlying application executed by {\tt CARE} is abstracted away. 
Therefore, properties related to the underlying application are not verifiable under the current abstraction. 
 An example of property not verifiable under the current abstraction could be to show that the returned payload of an action (e.g., an {\tt Integer} object) is always non-negative, by reasoning on the code of the corresponding method returning the value.
This kind of properties cannot be verified under the current abstraction. 
In this case, a theorem prover like KeY~\cite{DBLP:series/lncs/10001} would be more suited than {\sc Uppaal}. 

\subsection{Traceability} 
\begin{figure}[th]
    \centering
    \includegraphics[width=\columnwidth]{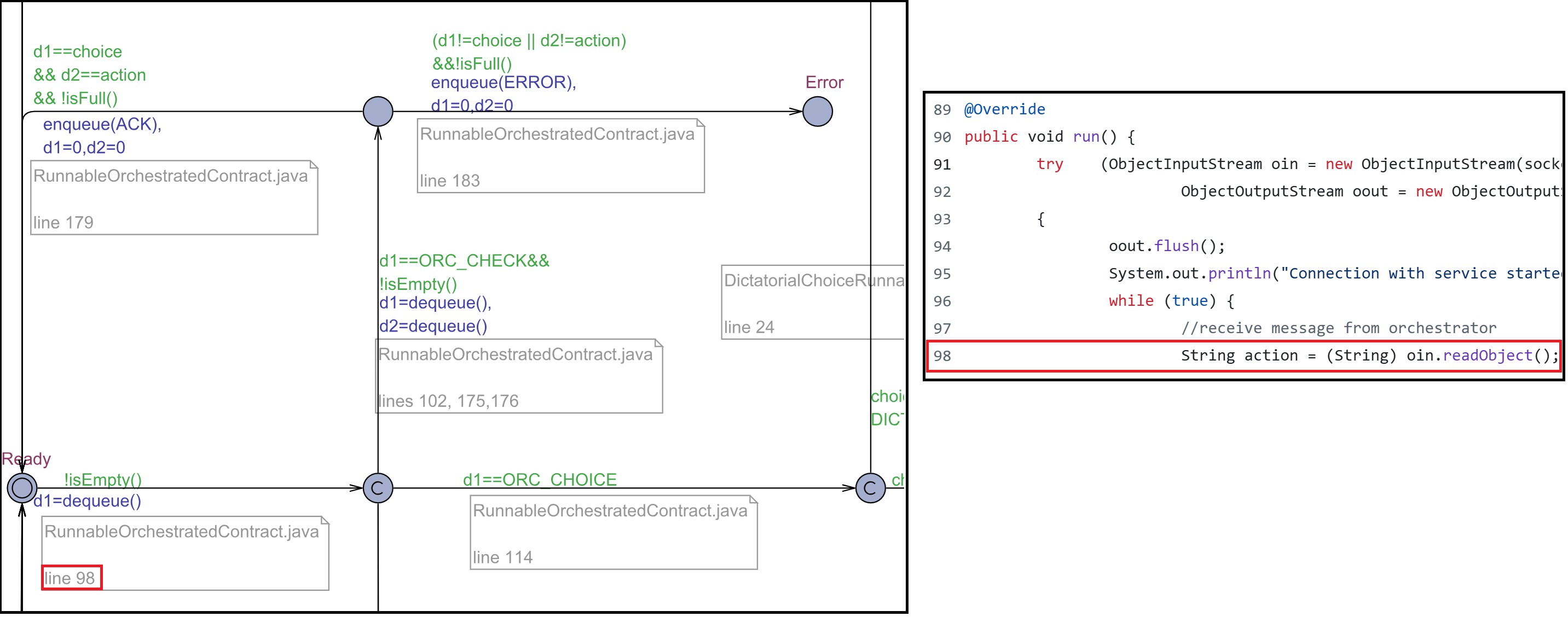}
    \caption{A fragment of the {\tt RunnableOrchestratedContract} automaton equipped with traceability information (left) and a fragment of the class {\tt RunnableOrchestratedContract.java} (right). The transition from state {\tt Ready} is traced back to line 98 of the source code of the respective class. }
    \label{fig:traceability}
\end{figure}

This paper adopts a lightweight approach to integrate the formal model with the concrete implementation, using the facilities provided by {\sc Uppaal}.
Traceability and model-based testing have been  used to ascertain the adequacy of the model with respect to the implementation.  
Traceability involves connecting the items of the abstract model with the lines of source code that are being abstractly represented. As stated previously, some aspects of the application are abstracted away. Therefore, only the relevant source code lines are connected with the abstract model.
In~\cite{UppaalModels} each transition of the model is traced back to the corresponding source code instructions using comments in the model. 
Traceability information is related to a specific version of the source code\footnote{\url{https://github.com/contractautomataproject/CARE/releases/tag/v1.0.1}}. 
For example, Figure~\ref{fig:traceability} shows a portion of the automaton modelling the {\tt RunnableOrchestratedContract} (discussed in details in the next section). 
Specifically, from state {\tt Ready} a transition is executed which is traced back to line~98 of the class {\tt RunnableOrchestratedContract}, also showed in Figure~\ref{fig:traceability}. 
Indeed, at line~98 the first read operation from the {\tt ObjectInputStream}, reading from the TCP/IP socket, is performed by the service. This is abstracted in the model by the operation {\tt dequeue()} of the automaton. 

Testing is used to show the adherence of the model to the actual implementation, and will be discussed in Section~\ref{sect:testing}. 

\paragraph{Remark}
We highlight the advantages of employing graphical diagrams. When it comes to the implementation phase, developers typically work with the source code, which currently consists of 770 lines of code in the case of {\tt CARE}.
On the other hand, during the modelling phase with {\sc Uppaal}, designers graphically edit automata. The automata depicted in Figure~\ref{fig:runnableorchestration} and Figure~\ref{fig:runnableorchestratedcontract} 
 succinctly and accurately specify the interaction logic of {\tt CARE}.

\section{Formal model}\label{sect:formalmodel}

\begin{figure}[t]
    \centering
    \includegraphics[width=0.7\columnwidth]{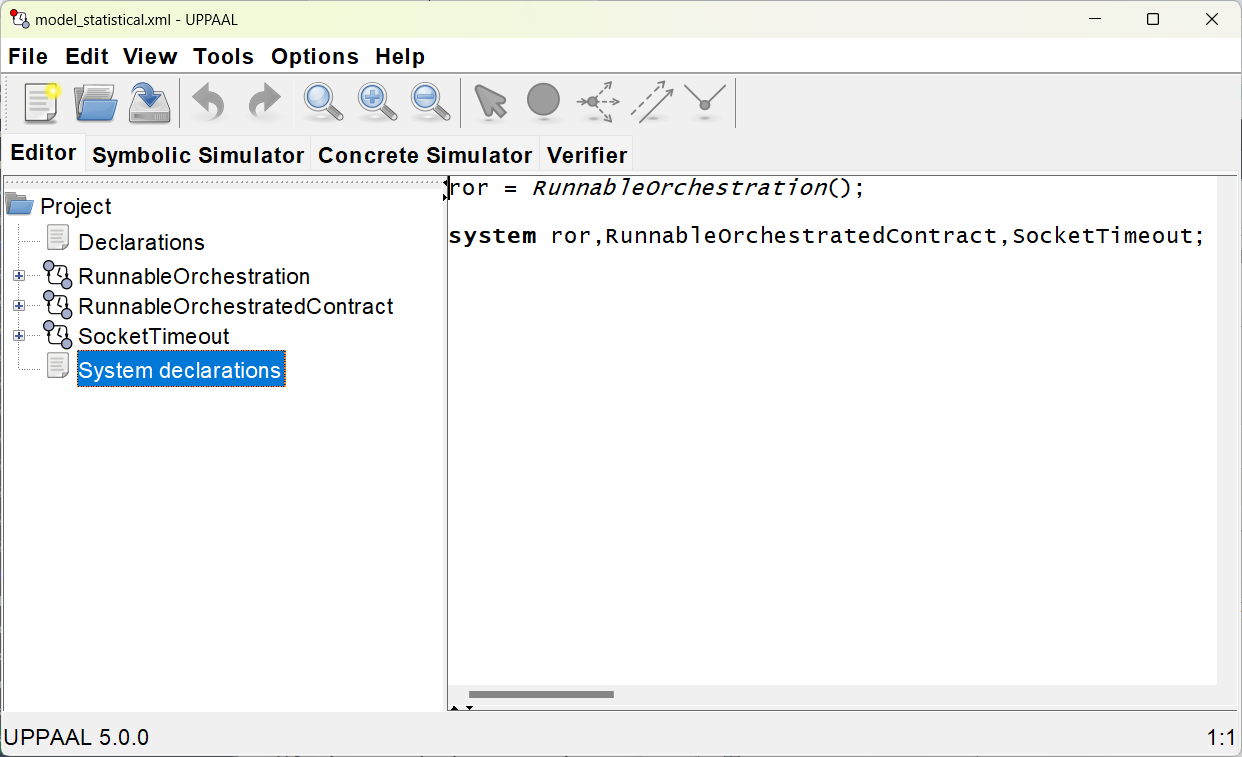}
    \caption{A snapshot of the {\sc Uppaal} model showing the system declarations}
    \label{fig:system}
\end{figure}

This section describes the formal model of {\tt CARE}. All models used in this paper together with the evaluated formulas are available in~\cite{UppaalModels}. 
The network of automata is composed of a {\tt RunnableOrchestration} automaton modelling the orchestrator, and each service is modeled by two automata:  {\tt RunnableOrchestratedContract} and {\tt SocketTimeout}. 
Figure~\ref{fig:system} shows the system declarations of the model. In particular, there is a single instance of  {\tt RunnableOrchestration} whilst there are {\tt N} instance of {\tt RunnableOrchestratedContract} and {\tt SocketTimeout}. 
Indeed, both {\tt RunnableOrchestratedContract} and {\tt SocketTimeout} templates take as parameter {\tt id\_t}, which is 
a special global declaration of {\sc Uppaal} (namely, {\tt typedef int[0,N-1] id\_t}) to instruct the model to produce {\tt N} replicas of these template, where as stated above {\tt N} is also a global constant of the model. 

\begin{figure}[t]
    \centering
    \includegraphics[width=0.8\columnwidth]{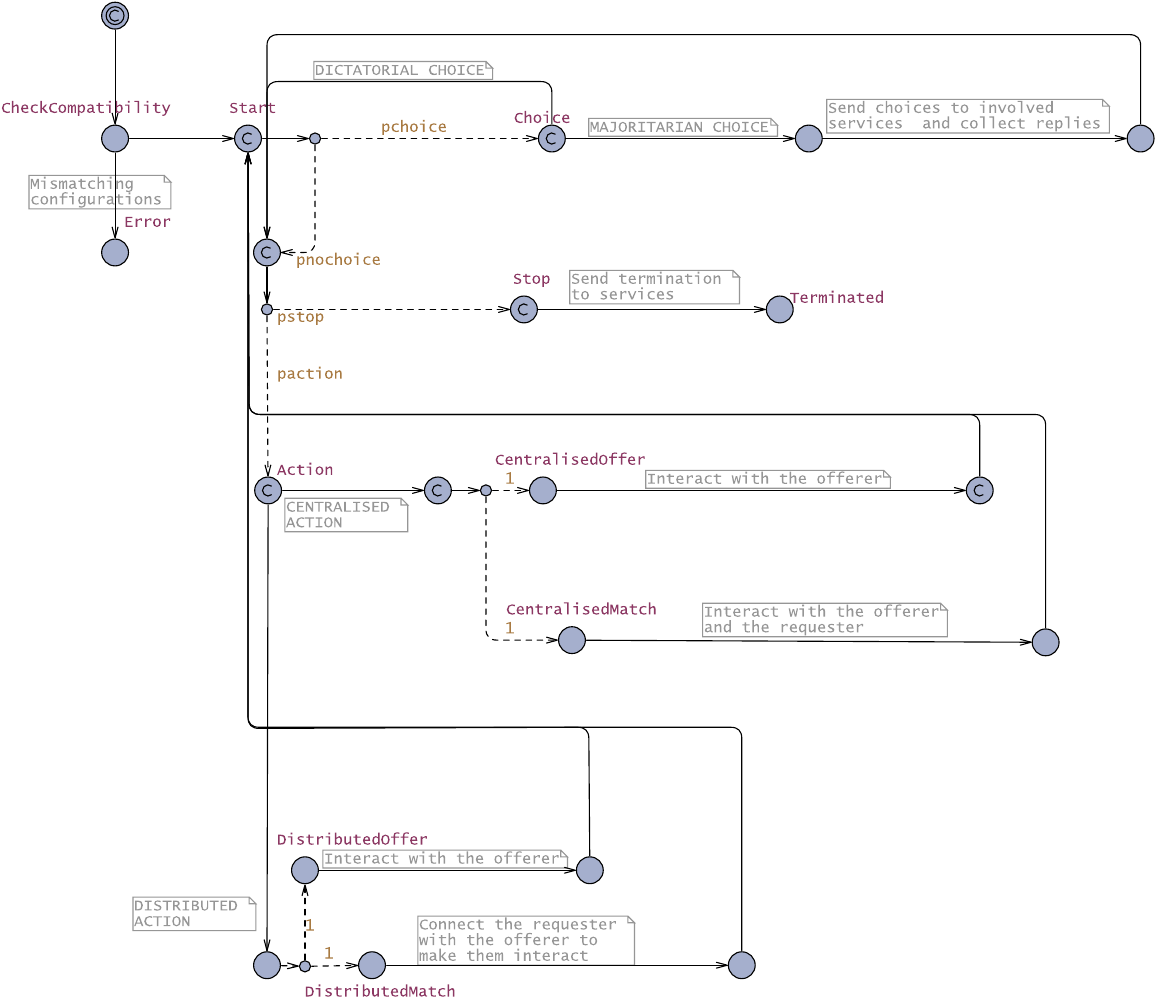}
    \caption{A simplified schema of the {\tt RunnableOrchestration} automaton}
    \label{fig:simplified}
\end{figure}

Figure~\ref{fig:runnableorchestration} displays the template automaton for the {\tt RunnableOrchestration} (i.e., the orchestrator), while the template automaton for the {\tt RunnableOrchestratedContract} (i.e., the service) has as parameter the {\tt id} of the service (of type {\tt id\_t}) and is depicted in Figure~\ref{fig:runnableorchestratedcontract}. Recall that, for readability, the {\tt Timeout} location is not displayed (see Section~\ref{sect:modellingtcpipsockets}). 
The behaviour according to the given configuration  of action and choice is modelled inside each automaton.
We anticipate that we will use variants of these two automata in Section~\ref{sect:analysis}, in order to perform different analyses. 
In particular, in Figure~\ref{fig:runnableorchestration}  
the configurations of choice and action (for both the services and the orchestrator) are instantiated non-deterministically by the transition outgoing the initial state (see Section~\ref{sect:descriptionofautomata}). Note that, in this case, the variables {\tt choice} and {\tt action} are not parameters of the templates. 
Another version of the model is also available, where the configuration options, instead of being selected non-deterministically, are fixed as parameters of the templates.
Note that in {\sc Uppaal} templates cannot be primitively instantiated non-deterministically.
The repository~\cite{UppaalModels} contains the model equipped with traceability information, i.e., each transition of the model has a comment describing the class and the lines of the source code that correspond to the specific behaviour of the transition. 
The models used for generating tests, together with the generated tests, are also available in~\cite{UppaalModels}.

%

%
The {\sc Uppaal} model is composed of a list of global declarations, the two automata (with their local declarations) and the system set-up (i.e., the instantiation of the automata). 
Global declarations include the number of services {\tt N}, the size of the buffers, the timeout threshold, the rates of the exponential distributions, two variables {\tt action} and {\tt choice} storing the corresponding configuration for all automata, and the communication buffers.
Constants are also defined globally and their identifiers are in capital letters.
Some names of locations are displayed in the automata for readability. 
Labels of transitions contain guards and effects and in a few cases also probabilistic selections. 

\subsection{Description of the automata}\label{sect:descriptionofautomata} 
Before discussing the two automata, we describe a simplified template, depicted in Figure~\ref{fig:simplified}.

\subsubsection{Simplified template}
Note that Figure~\ref{fig:simplified} is not a {\sc Uppaal} automaton.
The initial state is depicted with a double circle.
The first activity consists in checking whether all services and the orchestrator have the same configuration ({\tt CheckCompability}). If this check is successful, the orchestration starts.
From the location {\tt Start}, the orchestrator internally decides (based on the orchestration) whether to perform a choice or an action. The choice of termination is modelled as a third alternative.
Probability weights are used to model the probability of performing a choice, an action or to terminate ({\tt pchoice}, {\tt pstop}, {\tt paction}).
As stated previously, after a choice is completed, the orchestrator moves to a state where only an action or termination can be chosen. After an action is completed, the orchestrator returns to the {\tt Start} location.
The choice can be either dictatorial or majoritarian, according to the assigned configuration.
In the majoritarian choice, the orchestrator interacts with the services to collect their choice.
Similarly, the configuration of action can be either centralised or distributed.
In both cases, an offer label or a match label will be executed. As stated in Section~\ref{sect:background}, in a distributed match the orchestrator makes the two services aware of each other so that they can interact and execute the match.
In case of termination, the orchestrator sends the termination message to all services and terminates.

\subsubsection{Detailed description}
\begin{figure}[t]
    \includegraphics[width=1\columnwidth]{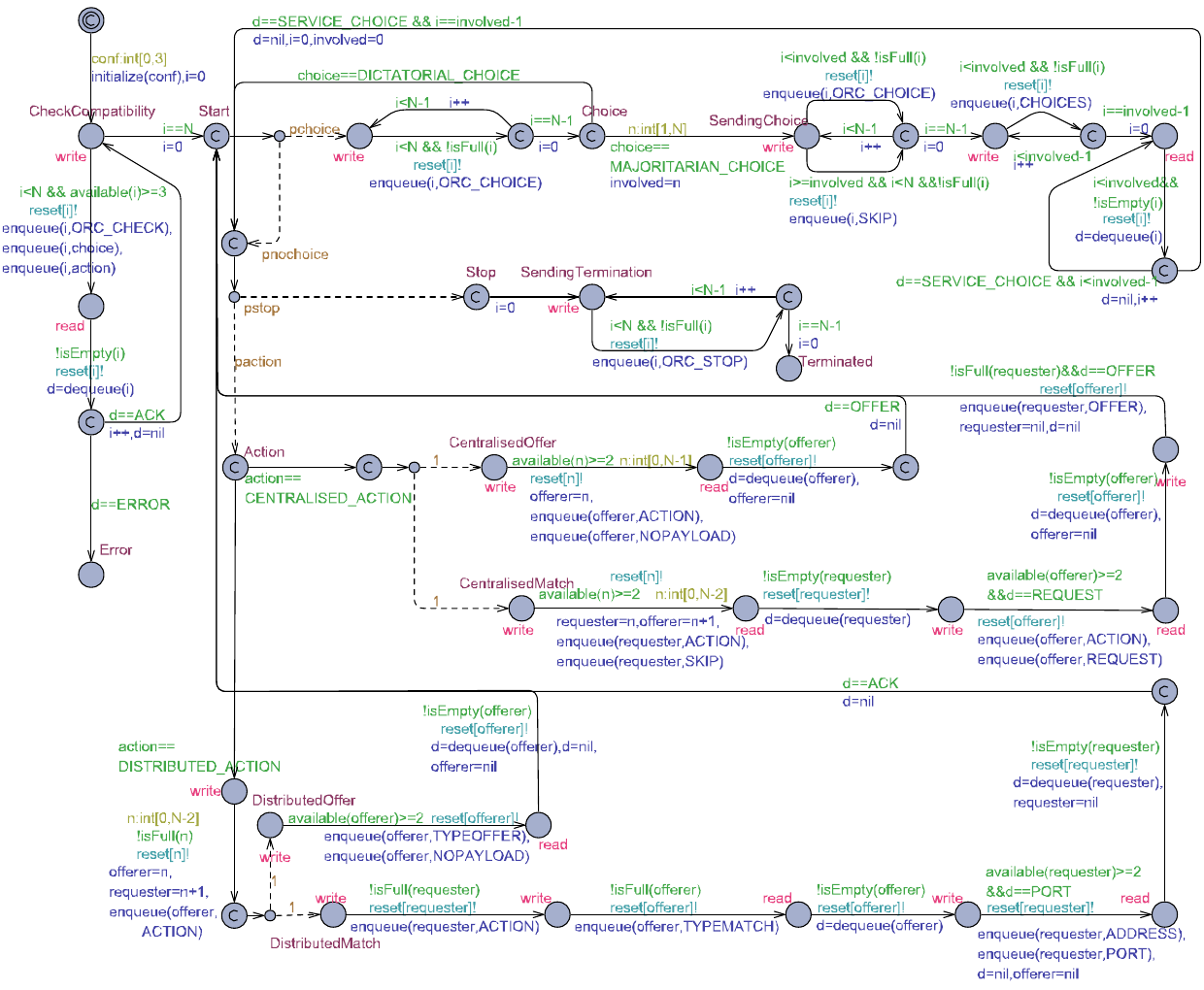}
    \vspace*{-\baselineskip}\caption{The {\tt RunnableOrchestration} {\sc Uppaal} template}
    \label{fig:runnableorchestration}
\end{figure}

\begin{figure}
    \includegraphics[width=\columnwidth]{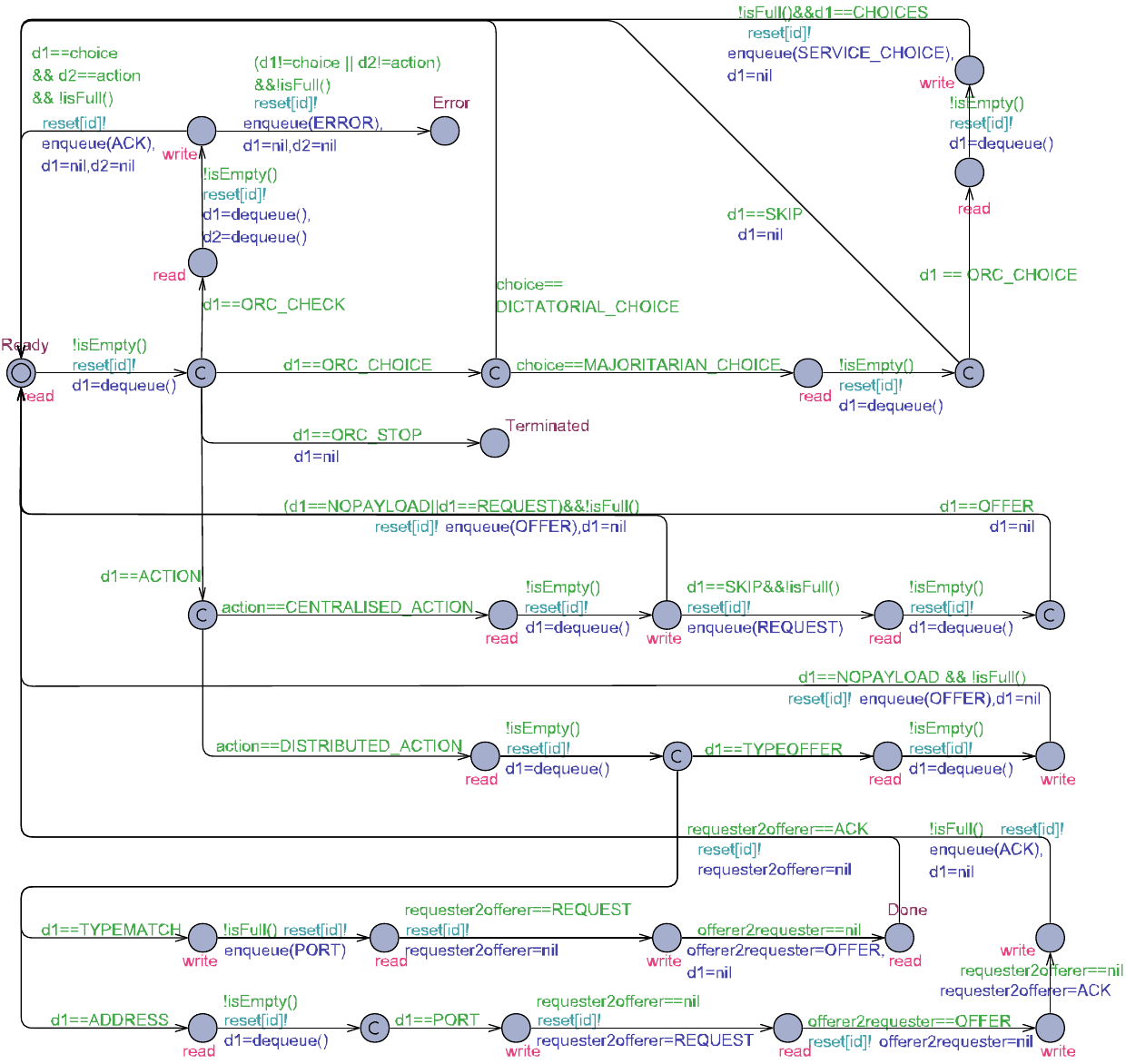}
    \vspace*{-\baselineskip}\caption{The {\tt RunnableOrchestratedContract} {\sc Uppaal} template 
     }
    \label{fig:runnableorchestratedcontract}
\end{figure}

We now briefly describe Figure~\ref{fig:runnableorchestration} and Figure~\ref{fig:runnableorchestratedcontract}. 
In the first transition of the orchestrator, one of four options encoding the combinations of choice and action is selected  non-deterministically. The function {\tt initialize(conf)} instantiates accordingly the {\tt choice} and {\tt action} global variables such that all have the same configuration, and also initialises the communication buffers. 

From the location {\tt CheckCompatibility} of the orchestrator a loop is executed to communicate with all services to check if all have the same configuration. 
The orchestrator sends the message {\tt ORC\_CHECK} and its configuration (action and choice).  If a service has the same configuration an {\tt ACK} is sent, or an {\tt ERROR} message otherwise.  
In case of no errors, the orchestration starts. 
As stated above, from  the  location {\tt Start} the orchestrator internally decides whether to perform a choice, an action or terminate. 


In the case of termination, the orchestrator sends to all services an {\tt ORC\_STOP} message and the orchestration terminates.

\paragraph{Choice}
In the case of a choice, firstly all services receive the message {\tt ORC\_CHOICE}. If the choice is dictatorial, the orchestrator decides autonomously, and no further interactions are necessary. Otherwise, in the case of a majoritarian choice, the services involved in the choice are selected non-deterministically, and a message {\tt ORC\_CHOICE} is sent to them. Concretely, the involved services are  those who perform an action in one of the outgoing transitions from the current state of the orchestration. The services that are not involved will receive a {\tt SKIP} message. 
The involved services then receive from the orchestrator the available choices (i.e., the forward star of the current state). 
These concrete choices are abstracted by the constant {\tt CHOICES}. 
Each involved service now replies with its choice, abstracted as a message {\tt SERVICE\_CHOICE}. 
After all involved services have voted, the orchestrator will decide accordingly.

\paragraph{Action}
In case of an action, the orchestration behaves differently depending on whether the configuration is centralised or distributed. 
In both cases, a probabilistic choice is made on whether the transition is an offer or a match. After that, the orchestrator picks non-deterministically  an offerer and a requester (only in the case of a match transition) with consecutive identifiers (recall that their IDs are immaterial).
In the case of a centralised offer action, the offerer  receives from the orchestrator the invocation of the  action, abstracted by the constant {\tt ACTION}. 
At this point the offerer is still not aware of whether it is  involved in an offer or a match transition. 
The reception of the  {\tt NOPAYLOAD} message (i.e., there is no payload from the requester) disambiguates  the offerer who replies with a payload  abstracted here by the constant {\tt OFFER}.

In a centralised match the requester receives the action invocation, abstracted by the constant {\tt ACTION}, and a {\tt SKIP} command (i.e., the offer has not yet been generated).
Note that in the implementation the service is informed of being an offerer or a requester upon receiving the action.  Since the concrete action is abstracted, in the model this disambiguation occurs when receiving either the message {\tt SKIP} (i.e., a requester) or the messages {\tt NOPAYLOAD} (i.e., an offerer in an offer transition) or {\tt REQUEST} (i.e., an offerer in a match transition).
The requester replies with the message {\tt REQUEST} (the concrete payload is abstracted away) that is forwarded by the orchestrator to the offerer, who receives in sequence the messages {\tt ACTION} and {\tt REQUEST}. 
Similarly to the previous case, the offerer sends to the orchestrator  its offer payload (now based on the payload of the requester), abstracted again by the constant {\tt OFFER}.

Concerning the distributed configuration, in case of an offer or match first the offerer receives the {\tt ACTION} command from the orchestrator. In case of an offer, a {\tt TYPEOFFER} message is received by the offerer followed by a {\tt NOPAYLOAD} message. The offerer replies with an {\tt OFFER} message. 
In case of a distributed match,  the {\tt ACTION} message is also sent to the requester and the {\tt TYPEMATCH} message is sent to the offerer. The offerer opens a fresh  port and communicates this  port (abstracted by the constant {\tt PORT}) to the orchestrator. The offerer waits for a connection from the requester.  The orchestrator communicates to the requester (who was waiting after receiving the {\tt ACTION} command) the address (constant {\tt ADDRESS}) and {\tt PORT} of the offerer. 
Now the requester and the offerer can interact without the orchestrator. 
The interactions between any two services in the orchestration all use two (one-position) buffers, implemented by the variables {\tt requester2offerer} and {\tt offerer2requester}. 
Indeed, it is never the case that  some service interferes in a match in which it is not involved, and using different buffers for each pair of services would unnecessarily increase the state space. 
Initially, the requester sends its {\tt REQUEST} payload to the offerer. The offerer replies to the requester with its payload {\tt OFFER} (based on the request) and reaches location {\tt Done}. 
The requester receives the payload and sends an {\tt ACK} to the offerer and the orchestrator (who both terminate the execution of the match transition and returns to the {\tt Start} and {\tt Ready} state, respectively).

\section{Analysis}\label{sect:analysis}

We now describe the analyses we performed on the model. 
The modelling activity led to some issues in both the implementation and the model, which were all fixed (we remark that an already existing application was modelled). 
Other formal checks on the model were performed to ensure the properties described in this section (e.g., absence of deadlocks, absence of orphan messages). 
We have used {\sc Uppaal} version 4.1.26-1, released in February 2022.
Note that a more recent version of {\sc Uppaal} is available, which integrates the previously separate tool {\sc Uppaal} Stratego.
We opted to continue using the version we initially started working with, since the capabilities of {\sc Uppaal} Stratego are not used in this paper.

\subsection{\bf Validation through modelling} 
The first validation was performed during the modelling phase. 
Indeed, 
formal modelling requires an accurate analysis and review of the source code. 
Interactive simulation is used during modelling to animate and analyse the portion of the model designed so far, similarly to how the source code is debugged interactively (e.g., by choosing the next step). %
We note that in many model-based engineering tools,   behavioural models (e.g., state charts) are validated by only relying on graphical interactive simulations~\cite{fmics/BasileMF23}.
Issues can be detected during this phase in particular if the source code has not been thoroughly tested, as was the case for {\tt CARE}. 

We report an issue detected in the source code during the modelling of the automaton in Figure~\ref{fig:runnableorchestration},
and more specifically during the modelling of the loop in which the orchestrator is reading the {\tt SERVICE\_CHOICE} messages sent by the involved services. 
In the implementation, the orchestrator was waiting for a choice from all services, also comprehending those who received a {\tt SKIP} message. 
This means that a deadlock could occur in case there was a service not involved in a choice. 
Initially, this issue was undetected because the tests had all services involved in all choices. It was fixed thanks to the activity described in this paper. 

On a side note, thorough testing is generally more time-consuming than designing a formal  model similar to the one in this paper.   
 This is a further benefit derived from the usage of formal methods. 
 For example, the library implementing contract automata operations ({\tt CATLib})~\cite{BB22OSP} has been tested up to 100\%  coverage of all lines and branches, also using mutation testing~\cite{BB22OSP}. In {\tt CATLib}, the lines of code of the tests are more than three times those of its  source code~\cite{BB22OSP}. 
 A similar effort for {\tt CARE} is more demanding than the one needed for designing the models in Figures~\ref{fig:runnableorchestration} and~\ref{fig:runnableorchestratedcontract}. 
{\sc Uppaal} has been used to automatically generate tests from the formal model.

\subsection{Formal Verification}
We discuss the formal verification performed on the model, encompassing both exhaustive and statistical model checking.
In the initial phase, we employ a model variant that guarantees consistent configurations between the services and the orchestrator. In this variant, the selection of the {\tt choice} and {\tt action} configuration is performed non-deterministically by the first transition of {\tt RunnableOrchestration} (variable {\tt conf}, see Figure~\ref{fig:runnableorchestration}), and it is always consistent. 
Subsequently, we move to another variant where the configuration of  {\tt choice} and {\tt action} are treated as parameters. In this case, we formally prove that when configurations do not match, an error state is reached.

\noindent
{\it Performances.} Statistical model checking has been used to scale to larger systems, and the verification  has been performed in a few seconds on a standard laptop. 
Conversely, exhaustive model checking necessitated more resources and has been limited to smaller parameter setup generating  hundreds of millions of states (see below).
In this case, the verification has  been performed on a machine with Intel(R) Core(TM) i9-9900K CPU @ 3.60\,GHz  equipped with 32\,GB of RAM. {\sc Uppaal} has been configured for  maximising the state space optimisation and reusing the generated state space. 
Logs of the experiments are available in~\cite{UppaalModels}. 
The summaries of the analysis performed with statistical (resp., exhaustive) model checking are in Table~\ref{tab:statistical} (resp., Table~\ref{tab:standard}).

\begin{table}[t]
\centering
\resizebox{0.99\columnwidth}{!}{
\begin{tabular}{lp{6cm}p{4cm}p{2.2cm}p{1.9cm}}

\textbf{Task} & \textbf{Property} & \textbf{Parameters' tuning} & \textbf{Interval} & \textbf{Confidence}\\ \toprule
Buffer size          &   Probability of filling a buffer & {\bf queueSize} set to 5 & $[0,0.00996915]$              & 0.95         \\ \midrule
Timeout         &   Probability of timeout   & {\bf timeout} set to 15 & $[0,0.00996915]$   & 0.95         \\ \midrule
Delays         &   Probability of termination & {\bf write} and {\bf read} set to 5 & $[0.990031,1]$    & 0.95         \\ \midrule
Orphan messages & Probability of terminating with non-empty buffers & - & $[0,0.00999882]$  & 0.995 \\ \midrule
Interference    & Probability of a service interfering on a match between two other services & - & $[0,0.00999882]$
 & 0.995 \\
\end{tabular}
} 
\caption{A summary of the properties analysed with statistical model checking }
\label{tab:statistical}
\end{table}

\begin{table}[t] 
\centering
\resizebox{0.99\columnwidth}{!}{
\begin{tabular}{lp{6cm}p{1.2cm}p{1.2cm}p{1.5cm}p{1.5cm}}
\textbf{Task}             & \textbf{Property}                                                                                   &  \textbf{Conf.} & \textbf{Time} & \textbf{Memory} & \textbf{State space} \\ \toprule
\multirow{2}{*}{Termination} & \multirow{2}{6cm}{If orchestrator terminates then all terminate or a timeout occurs}     &  c1                & 3 m.                 & 1.5 GB         & 52 M st.      \\ \cmidrule(r){3-6}
                          &                                                                                                     &                                                    c2                & 30 m.               & 12 GB    & 426 M st.              \\ \midrule
\multirow{2}{*}{Deadlocks}   & \multirow{2}{6cm}{Absence of deadlocks (no enabled transitions, timeout, or termination)}          &  c1                & 3 m.                  & 1.5 GB     & 25 M st.            \\ \cmidrule(r){3-6}
                          &                                                                                                     &                                                    c2                & 40 m.                & 13 GB     & 189 M st.             \\ \midrule
\multirow{2}{*}{Orphan Messages} & \multirow{2}{6cm}{Upon termination, no messages left in buffers}                              &  c1                & 2 m.                   & 1.5 GB    & 25 M st.             \\ \cmidrule(r){3-6}
                          &                                                                                                     &                                                    c2                & 27 m.                 & 12.5 GB        & 189 M st.       \\ \midrule
\multirow{2}{*}{Dummy Execution} & \multirow{2}{6cm}{No dummy execution where no interaction occurs}                              &  c1                & 3 sec.                & 1.4 GB   & 24 st.               \\ \cmidrule(r){3-6}
                          &                                                                                                     &                                                    c2                & 24 s.                & 10 GB     & 28 st.            \\ \midrule
Compatibility Check      & Mismatch leads to \texttt{Error} location                                                           &   Special config.   &   1 ms.       & 49 MB   & 79 st.              \\
\end{tabular}
}
\caption{A summary of the properties analysed with exhaustive model checking that are satisfied by the model }
\label{tab:standard}
\end{table}

\subsection{Parameters Tuning}
The verification process involves employing specific parameter setup within the model.  
This encompasses various aspects, such as setting the  delays in reading and writing, setting socket timeout thresholds, adjusting buffer sizes, assigning probability weights, and determining the number of services involved (i.e., the instantiations of the {\tt RunnableOrchestratedContract} template). 
For deriving the desired set-up of parameters, we employ statistical model checking. If not stated otherwise, the parameters $\alpha$ and $\epsilon$ of the statistical model checker are set to $0.05$ and $0.005$, respectively (see Section~\ref{sect:background}). 

In this section, when reporting the evaluation of a probabilistic formula, instead of reporting an interval containing zero (e.g., $[0,0.00996915]$), sometimes, for conciseness, we will say that the property has a probability near zero. To be more precise, we mean that all runs do not satisfy the property, thus the probability lies within the interval $[0,0+\epsilon]$ with probability $1-\alpha$. We remind readers that, unlike probabilistic model checking, in statistical model checking, the unknown probability value is estimated to lie within an interval with a given confidence.

It is crucial to note that we do not employ statistical model checking to determine values (such as buffers size) for use in the concrete  implementation. 
Instead, these quantified values are used solely within the model.  
For instance, in the actual implementation, the size of Java TCP/IP socket buffers is fixed (more below).
Our objective is to employ parameter setup that ensure realistic modelling and improve the performances of the exhaustive model checking. 
Realistic modelling entails accurately representing the behaviour of the real system, by reducing the probability of filling the buffers, timeouts, or excessive communication delays.  
Failure to maintain these conditions could potentially invalidate the results of the formal verification.
Improving performances, on the other hand, entails reducing the state space of the model. 

\begin{figure}[t]
    \centering
\includegraphics[width=0.3\textwidth]{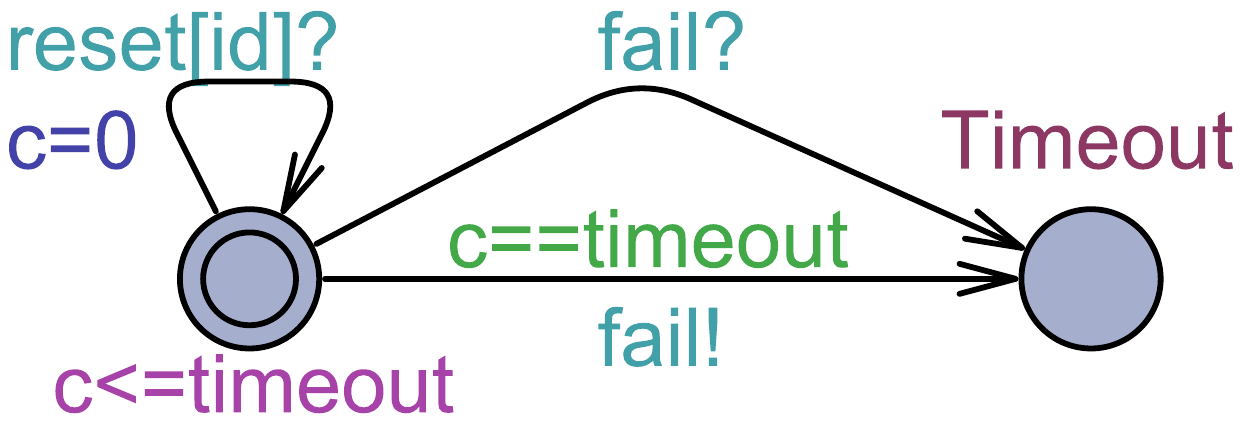}
\qquad
\includegraphics[width=0.49\textwidth]{clientxservice.pdf}
    \vspace*{-\baselineskip}\caption{On the left side, the {\tt SocketTimeout}  template automaton (one such automaton is istantiated for each service). On the right side, the orchestration automaton taken from the composition service example in~\cite{DBLP:conf/fm/BasileB23}}
    \label{fig:sockettimeoutandorchestration}
\end{figure}

\medskip
\noindent
{\it Probabilities weights.}
The values of the {\it probabilities weights} 
{\tt pstop}, {\tt pchoice}, {\tt pnochoice}, and {\tt paction} (see Figure~\ref{fig:runnableorchestration}) can be tuned based on  average values extracted from   the orchestrations subject of the analysis. Indeed, as stated in Section~\ref{sect:methodology},  the underlying orchestration automaton is abstracted away. 
Note that, e.g., in location {\tt Start} the probability of performing a choice  is $\frac{\tt pchoice}{\tt pchoice+pnochoice}$ ({\tt pchoice} is the probability weight of performing a choice and {\tt pnochoice} is the probability weight of not performing a choice). 
For example, the orchestration in Figure~\ref{fig:sockettimeoutandorchestration} (right) can be modelled by tuning the probabilities to {\tt pstop=25}, {\tt pchoice=1}, {\tt pnochoice=0}, and {\tt paction=75}.

For readability, in the reminder of this section we will use in the formulae the abbreviations {\tt ror} for the instantiation of the {\tt RunnableOrchestration} template (i.e., the orchestrator), and {\tt ROC(i)} for the i-th instantiation of the {\tt RunnableOrchestratedContract} template (i.e., a service).

\medskip
\noindent
{\it Buffer size.}
Next, we address the {\it buffer size} (denoted as variable {\tt queueSize}).     
It is important to note that the buffers in the model are represented by global arrays utilized by the automata for enqueuing and dequeuing values. These global arrays serve as models of the actual buffers found in Java TCP/IP sockets, where the default size is typically 8 KB.
Our objective is to prevent unnecessary growth in the model's state space while ensuring a low probability of the buffers filling up, similar to the behaviour observed in the real application.
The formula:
\begin{center}
{\scriptsize\tt
E[<=500; 10000](max: sum(i:int[0,N-1]) (sum(j:int[0,queueSize-1])(orc2services[i][j]!=nil)))
}
\end{center}

\noindent
computes, using 10000 simulations of 500 time units, the expected maximum number of non-empty positions of the buffers used by the orchestrator to send messages to the services (called {\tt orc2services}). 
For all statistical evaluations, we set the number of services to $10$.  
With buffer size set to $10$, this formula evaluates to $4.5 \pm 0.019$. 
This indicates that, on average, the maximum number of utilized buffer positions is between 4 and 5. Consequently, in order to reduce the state space, it is safe to decrease the value of {\tt queueSize} to less than 10 for the exhaustive model checking phase.
This observation is further supported by the following formula:

\begin{center}
{\scriptsize\tt{Pr[<=500] (<>(exists (i:id\_t) ror.isFull(i)))}}
\end{center}
which measures the probability that, within 500 time units, one of the {\tt orc2services[i]}  arrays becomes full ({\tt ror} is the orchestrator automaton). In three separate experiments where {\tt queueSize} was varied between 3, 4, and 5, this formula yielded the respective evaluation intervals of $[0.990031,1]$, 
$ [0.00632077,0.0163207]$ (with  $\alpha\!=\!0.005$), 
and $[0,0.00996915]$. Based on these results, if not stated otherwise, {\tt queueSize} is  set to 5 for the subsequent experiments.

\medskip
\noindent
{\it Delays and timeout.}
Next, we consider the real-time behaviour of the model and focus on determining appropriate values for the {\it rates}  {\tt write} and {\tt read}, which represent the delays in writing to and reading from a buffer, respectively.
The average message delay in Java TCP/IP sockets is affected by multiple factors, such as network conditions and server load. Delays can be sampled using tools like {\tt tcpdump}, and the rate can be estimated by calculating the inverse of the sample mean.
Additionally, we consider the variable {\tt timeout}, which represents the timeout threshold.
Our aim is to achieve three objectives. First, we strive to maintain a low probability of encountering timeouts. 
Second, we seek to ensure a high probability of terminating within a specific timeframe, which we have set to be 500 time units, corresponding to the duration used in our experimental setup. 
Third,  we aim to keep the timeout threshold at a lower value in order to decrease the state space and facilitate model checking.

The formula:
\begin{center}
{\scriptsize\tt{Pr[<=500] (<> ror.Timeout)}}
\end{center}
 measures the likelihood of one or more service sockets experiencing a timeout (recall that in this case a failure signal is broadcasted and all automata enter their respective {\tt Timeout} location). 
The probability of all services and the orchestrator successfully concluding their operations
is evaluated with the formula:
\begin{center}
{\tt \scriptsize{
Pr[<=500](<>ror.Terminated\&\&(forall (i:id\_t) ROC(i).Terminated))
}}
\end{center}
  ({\tt ROC} is used as an abbreviation of {\tt RunnableOrchestratedContract}).
In different experiments  where we varied the  values of the pair (rate,timeout), specifically in $(5,14)$, $(4,15)$, $(5,15)$, respectively, the first formula (timeout probability) yielded the respective evaluation intervals
$ [3.06006e^{-06},0.00999494]$ (with   $\alpha\!=\!0.005$),
$[0.0433422,0.053341]$, and  
$[0,0.00996915]$, while the second formula (probability of termination) yielded %
$[0.990005,0.999997]$ (with   $\alpha\!=\!0.005$), 
$ [0.948625,0.958625]$ (with   $\alpha\!=\!0.005$), and
$[0.990031,1]$.
Therefore, to fulfill the aforementioned three objectives, we have set the values of {\tt write} and {\tt read} to 5, while the value of {\tt timeout} has been set to 15 for the subsequent experiments. Indeed, in the experiments we just mentioned, the value (5,15) for the pair (rate,timeout) yields the best values for both the probability of experiencing timeout and the probability of reaching termination. 

Concerning the {\it instances} of the templates, in all experiments there is one instance of the orchestrator template and either 4 or 5 instances of the service template.  
For the exhaustive model checking phase, we used two small parameter setup of (number of services, buffers size). 
The first is c1=(4,5), the second is c2=(5,3). 
In fact, the parameter setup (5,4) remained inconclusive in the experiments due to the need for generating billions of states and the inadequate memory capacity of the utilized machine. 
The verification of larger parameter setup necessitates either relying exclusively on statistical model checking or employing more powerful machines.

\subsection{Verification} 
Once the model's parameter setup is determined, our next step involves verifying additional formal properties. 

\medskip
\noindent
{\it Termination.}
We have already assessed the probability of non-termination and found it to be nearly zero. 
However, it may be worthwhile to conduct an exhaustive verification specifically for this property. 
The property that in all executions eventually all services and the orchestrator terminate is not valid. 
Indeed, as described in Section~\ref{sect:methodology}, the orchestration contract automaton is abstracted away and 
 at each iteration, a choice is performed to decide whether to terminate or not. 
 Hence, there exists an execution in which the orchestration never terminates. 
A milder property does hold: 
\begin{center}
{\tt \scriptsize{ror.Stop-->((ror.Terminated\&\&(forall(i:id\_t)ROC(i).Terminated))||\\(exists(i:id\_t)SocketTimeout(i).Timeout))
}}
\end{center}
%

\noindent  i.e., if the orchestrator starts to terminate then eventually all services and the orchestrator terminate. The formula {\scriptsize \tt p-->q} is a shortcut for {\scriptsize \tt A[](p imply A<>q)}.
Thus, this formula states that for all executions and for all states either the orchestrator is not in the location {\tt ror.Stop} or all executions passing through that location will eventually lead to a state where all services and the orchestrator have terminated or a timeout failure is experienced. 
The formula holds in both parameter setup c1 and c2.
The first (resp. second) parameter setup required roughly 3 (resp. 30) minutes, 1.5 (resp. 12) Giga of memory, and explored  52 (resp. 426) million states.

\medskip
\noindent
{\it Absence of deadlocks.} 
The likelihood of non-termination being very low also implies an almost negligible probability of encountering deadlocks. While it may be of interest to exhaustively prove the absence of deadlocks, the previous formula is insufficient for this purpose.
Hence, to prove that there is no deadlock, we perform exhaustive model checking of the formula:

\begin{center}
{ \tt \scriptsize {
A[](not deadlock || (exists(i:id\_t) SocketTimeout(i).Timeout) ||\\ 
 (ror.Terminated  \&\& (forall (i:id\_t) ROC(i).Terminated)))}}
\end{center}

\noindent 
The formula states that for all executions and for all states of the composed system, either there is always at least one enabled transition or either a timeout failure has been experienced or all services and the orchestrator are in the {\tt Terminated} location. In the formula, {\tt not deadlock} is a special predicate provided by {\sc Uppaal}. 
As expected, this property is satisfied in both parameter setup.
The first (resp. second) parameter setup required roughly 3 (resp. 40) minutes, 1.5 (resp. 13) Giga of memory, and explored  25 (resp. 189) million states.
This also proves that in a correct configuration (of action and choice) the {\tt Error} location is never reached, because this would result in a deadlock and the above property would not be satisfied. 
This property is also satisfied when the size of the buffers is $3$. 
However, if we further reduce the size, 
then a deadlock occurs. 
This is because from location {\tt CheckCompatibility} the orchestrator requires to insert three messages in the buffer of the receiver in one step. 
By dividing these three send operations in three non-committed transitions it is possible to further reduce the buffer size.

We report a modelling issue detected during model checking the above formula.
In an earlier version, it was assumed that the socket mode was non-blocking (i.e., sending to a recipient with a full buffer would cause an error).
This was modelled by making committed ({\tt C}) all source states of transitions with sending operations.
In this way, if the buffer of the receiver would not have enough free space then an attempt to send to the receiver would cause a deadlock. 
In fact, in this earlier version of the model the above formula (absence of deadlocks) was not satisfied, for any possible size of the buffer. 
The counterexample trace of the model checker helped to understand and eventually fix this issue. 
Basically, in the model configured with a majoritarian choice there exists a loop in which the orchestrator alternates choices and actions, and  enqueues a sequence of {\tt ORC\_CHOICE} and {\tt SKIP} messages to a service that never consumes them and is never involved in neither  choices  nor actions, thus eventually filling its buffer and deadlocking. A similar issue also exists in the case of a dictatorial choice. 

Note that this kind of problems are 
hard to detect without model checking. 
Indeed, the counterexample trace was generated automatically, and the counterexample trace is composed of hundreds of  steps. 
Without model checking this would require to manually execute each step of this trace, and the longer the trace the less chance to discover it. 
After we detected this issue, 
we analysed the underlying Java TCP/IP socket semantics~\cite{JavaSocket} 
and fixed the model 
as described in Section~\ref{sect:methodology} (i.e., by modelling these 
sockets with default blocking mode). 
Another fix could be to include an ack after the reception of a {\tt SKIP} message (this would however require to modify also the implementation).

\medskip
\noindent
{\it Absence of orphan messages.} We now prove that upon termination of an orchestration no messages are left in any buffer, i.e., all messages are consumed. 
To expedite the verification process, we begin by conducting statistical model checking of the following property:
\begin{center}
{\scriptsize\tt{
Pr[<=500](<>!allEmpty()\&\&ror.Terminated\&\&(forall(i:id\_t)ROC(i).Terminated))
}}
\end{center}

\noindent
this property quantifies the probability of termination with at least one message remaining in any buffer. To verify whether all buffers are empty, we utilize the predicate \texttt{allEmpty()}. As expected, the probability is found to be nearly zero. We proceed with an exhaustive verification by employing the property:

\begin{center}
{\scriptsize\tt{
A[]((ror.Terminated \&\& (forall (i:id\_t) ROC(i).Terminated)) imply allEmpty())
}}
\end{center}

\noindent the above formula can be read as follows: in all states of all executions either all buffers are empty or someone has not terminated yet. 
The above property is valid in both parameter setup, as expected. 
The first (resp. second) parameter setup required roughly 2  (resp. 27) minutes, 1.5 (resp. 12.5) Giga of memory, and explored  25 (resp. 189) million states.
It is also possible to verify that there is no dummy execution in which the services and the orchestrator never interact (and thus all buffers are trivially empty). 
This can be verified with the formula:
\begin{center}
{\tt \scriptsize {E[] (allEmpty() \&\& !ror.Timeout)}} 
\end{center}
that, as stated above, checks if there exists an execution where all states have empty buffers (excluding the dummy execution scenario in which the timeout occurs at the beginning). As expected, this property is not valid in the model, for both parameter setup. 
The first (resp. second) parameter setup required roughly 3  (resp. 24) seconds, 1.4 (resp. 10) Giga of memory, and explored  24 (resp. 28) states.

\medskip
\noindent
{\it No interference.} When  discussing the distributed match action in Section~\ref{sect:formalmodel}, we stated that  it is never the case that  some service interferes in a match in which it is not involved. 
This guarantees that it is safe to use two one-position buffers for all communications between any two services involved in a match. 
To verify this, we perform statistical model checking of the following formula:
\begin{center}
{ \tt \scriptsize {
Pr[<=500](<>exists(i:id\_t)(i<N-1\&\&(ROC(i).d1==TYPEMATCH||ROC(i).d1==ADDRESS||ROC(i).d1==PORT)) \\
\&\&((ROC(i+1).d1==TYPEMATCH||ROC(i+1).d1==ADDRESS||ROC(i+1).d1==PORT))\&\& \\
(exists(j:id\_t)(j!=i\&\&j!=i+1\&\&(ROC(j).d1==TYPEMATCH||ROC(j).d1==ADDRESS||ROC(j).d1==PORT))))
}}
\end{center}
\noindent 
the formula measures the probability of reaching a state 
where one service (index~{\tt j}) is interfering on a match between two other services (indexes {\tt i} and {\tt i+1}). 
We recall that in the model two matching services have consecutive indexes. 
We detect a service to be involved in a distributed match when its temporary variable {\tt d1} has one of the three values ({\tt TYPEMATCH}, {\tt ADDRESS}, {\tt PORT}). 
The probability is found to be nearly zero.

We also include in the repository~\cite{UppaalModels} a version of the model where each pair of services has its own buffers. All results in this section also hold in that model.

\medskip
\noindent
{\it Compatibility check.} 
Next, we formally prove that if some service is not matching the configuration of the orchestrator, then the orchestration will not start and   an {\tt Error} location will always eventually be reached. 
Indeed, the possibility of mismatching configurations is allowed in the real system. 
However, in the model discussed in Section~\ref{sect:formalmodel} this scenario is not possible because, by construction, all services and the orchestrator 
share the same configuration. The configuration is selected non-deterministically. 
In this way, for each formula being verified all possible  configurations are checked automatically.

Only for this check  the model has been slightly modified  by adding two parameters {\tt action} and {\tt choice} to the templates and by updating accordingly the model.  
For this verification, the set-up of the system is of an orchestrator {\tt ror} and three services {\tt alice}, {\tt bob} and {\tt carl} and the size of each buffer is $4$: \\

\noindent\hfil 
{\tt \scriptsize 
{\parbox{0.84\columnwidth}{
ror = RunnableOrchestration(MAJORITARIAN\_CHOICE,DISTRIBUTED\_ACTION);\\
alice = RunnableOrchestratedContract(0,MAJORITARIAN\_CHOICE,DISTRIBUTED\_ACTION);\\
bob = RunnableOrchestratedContract(1,MAJORITARIAN\_CHOICE,DISTRIBUTED\_ACTION);\\
carl = RunnableOrchestratedContract(2,DICTATORIAL\_CHOICE,DISTRIBUTED\_ACTION);\\
ast = SocketTimeout(0);bst = SocketTimeout(1);cst = SocketTimeout(2);\\
system ror,alice,bob,carl,ast,bst,cst;
}}
}\par 
\phantom{..}\\

\indent Note that, differently from Figure~\ref{fig:system}, the configurations of choice and action are now  parameters assigned to each automaton. 
 This allows us to assign a mismatching configuration to verify that the {\tt Error} location will be reached.
Indeed, in the above set-up {\tt carl} has a different configuration. We use the formula:
\begin{center}
{\tt \scriptsize {A<>((ror.Error \&\& carl.Error)||ror.Timeout)}}
\end{center}
stating that in all executions eventually the orchestrator and the service with a wrong configuration reach an {\tt Error} location or a timeout is experienced.
Alternatively, the formula:
\begin{center}
{\tt \scriptsize {A[](!ror.Start)}}
\end{center}
states that in all executions the {\tt Start} location of the orchestrator is never traversed (i.e., the orchestration never starts).
Both properties are satisfied in this setup, thus verifying the correctness of the compatibility check. 
The first (resp. second) formula required visiting 19 (resp. 79) states. Both formulae used roughly 48 Megabytes of memory and a few milliseconds of CPU.
We have proved that in case of mismatching configurations, the orchestration will not start. 

\section{Testing} \label{sect:testing}
The model-based testing functionality of {\sc Uppaal} has been employed to generate tests that demonstrate the model's adherence to the actual implementation. 

%
 The abstract test cases generated by {\sc Uppaal} have been transformed into concrete JUnit tests for the source code. 
 {\sc Uppaal} provides various methods for test generation (see Section~\ref{sect:background}). 
 We utilized the generation from reachability queries (i.e., of the form {\tt E<>}). 
 In our models, these queries encode specific simulation traces that are relevant to the specific orchestration employed in the tests and are designed to cover all transitions of the model. 
  The specific simulation traces also guide the generation of the concrete test cases from the abstract test cases.
{\sc Uppaal} provides functionalities for offline test generation, such as achieving transition coverage or generating traces that reach a specified final state (reachability queries). 
However, {\sc Uppaal} does not primitively offer a mechanism within their query language to explicitly enforce a sequence of intermediate states or events that a test trace must follow. 
 To achieve this, two global variables, namely {\tt steps[]} and {\tt step}, are introduced in the models to encode the desired orchestration in the query. 
 Whenever a specific transition is executed, the variable {\tt steps[step]} is updated with a value encoding the transition being executed, and the counter {\tt step} is incremented. 
 For example (see Figure~\ref{fig:RunnableOrchestratedContractSteps}), whenever the transition reaching state {\tt Terminated} is executed, the value {\tt ORC\_STOP} is assigned to {\tt steps[step]}. As another example,  whenever a centralised match transition is executed by the orchestration, the value  {\tt CENTRALISED\_MATCH} is assigned to {\tt steps[step]}, encoding the execution of the request action in the match (see Section~\ref{sect:formalmodel}).  
 The length of the array {\tt steps[]} is equal to the desired number of transitions that the trace must pass through, and the variable {\tt step} is a pointer to the current position in the array. 
 As a small example, let {\tt alice} be an instatiation of the automaton in Figure~\ref{fig:RunnableOrchestratedContractSteps}. The query {\tt E<>(alice.steps[0]==ORC\_CHECK)} encodes a trace where the check operation is executed first.  
 In Section~\ref{sect:example}, we will provide a complete example showing how these additional variables are used in a query. 
 
 \subsection{Testing the Orchestrator}
Different variants of the model have been developed, each one equipped with specific test code to test one of the components. 
One version of the model is equipped with test code for testing the orchestrator (i.e., {\tt RunnableOrchestration}). 
In this case, during concrete testing, the orchestrator will be the actual system, whilst the tester will mimic the services. 
Therefore, the test code will be generated from the services model. 
In this version, the automaton template {\tt RunnableOrchestratedContract} contains test code in its transitions, whilst the automaton template {\tt RunnableOrchestration} does not contain test code. However, useful comments are inserted as test code in some transitions of {\tt RunnableOrchestration}  to guide the generation of concrete tests from abstract tests. 
This mainly involves printing the actual value during simulation of variables {\tt offerer}, {\tt requester} and {\tt i} used by the  {\tt RunnableOrchestration} as indices to identify the services the orchestrator is interacting with. 
In the generated tests, no timeout is experienced, therefore the {\tt SocketTimeoutAutomaton} is a dummy automaton receiving signals from {\tt RunnableOrchestratedContract} and doing nothing. Indeed, a timeout in the concrete system could be experienced due to a worsening of the physical network, which is outside the scope of the model-based testing performed in this paper to ascertain the adequacy.

\begin{figure}
    \centering
    \includegraphics[width=\columnwidth]{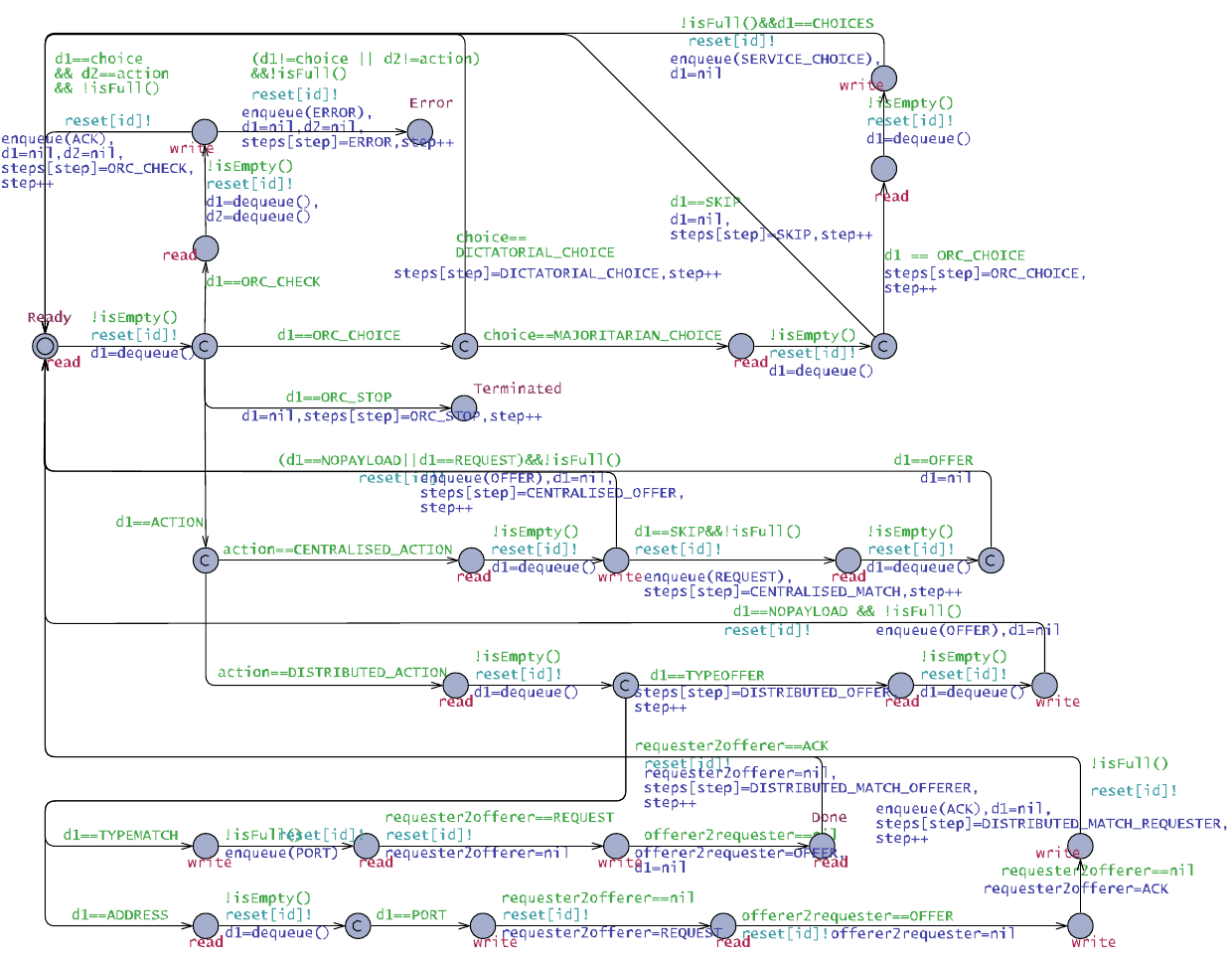}
    \caption{The {\tt RunnableOrchestratedContract} {\sc Uppaal} template used for testing the orchestrator }
    \label{fig:RunnableOrchestratedContractSteps}
\end{figure}

All transitions of {\tt RunnableOrchestratedContract} containing a {\tt dequeue()} operation in the effects will also contain as test code the instruction {\tt msg = (String) oin.readObject()}. 
Basically, the abstract operation of reading a message from the queue is concretised as a read operation from the variable {\tt oin} of class {\tt ObjectInputStream} of the Java TCP/IP socket of the service. 
Similarly, transitions containing an {\tt enqueue} effect will contain as test code instructions for writing into a socket. 
The information on the actual payload to be sent by the concrete services is abstracted away in the {\sc Uppaal} model. 
For example, consider the transition reaching state {\tt Ready} of {\tt RunnableOrchestratedContract} with effect {\tt enqueue(SERVICE\_CHOICE)} (see Figure~\ref{fig:RunnableOrchestratedContractSteps}, top right corner). In this case, {\tt SERVICE\_CHOICE} is a constant abstracting from the choice that the service will be expressing, which will be based at runtime on the transitions outgoing the current state of the contract automaton of the service, abstracted in the {\sc Uppaal} model. 
Accordingly, the test code generated from that transition is {\tt oout.writeObject(choice);oout.flush();}. 
In this case, {\tt choice} is a placeholder that needs to be replaced in the concrete test with the actual choice made at that point. Furthermore, {\tt oout} is the {\tt ObjectOutputStream} of the service's TCP/IP socket, on which a write operation is performed. 

Finally, some read operations and other transitions with guards will include assertions in their test code. These assertions will be concretised in the concrete tests by adding values that were abstracted away in the model and will be used by the JUnit tests to verify whether the tests are successful or not. 
For example, the transition of {\tt RunnableOrchestratedContract} reaching state {\tt Terminated} will contain in its test code the instruction {\tt assertEquals(msg,RunnableOrchestration.stop\_msg);}. 
In this case, it is tested whether the last received message is the constant identifying the termination of the current orchestration.

 \subsection{Testing the Services}

For testing the {\tt RunnableOrchestratedContract} services another version of the model is used.
The orchestrator will be mocked by the test, whilst the services will be the actual classes of {\tt CARE}.
The abstract test code will be generated through the automaton {\tt RunnableOrchestration}.
The {\sc Uppaal} automaton {\tt RunnableOrchestration} will also be decorated with the effects on the array {\tt steps[step]}, similar to Figure~\ref{fig:RunnableOrchestratedContractSteps}.
The generation of test code is similar to the case of testing the orchestrator, discussed previously.
The main difference is that the orchestrator manages a list of input/output streams, one for each service.
For example, when sending the message {\tt ORC\_CHOICE} to the services (see Figure~\ref{fig:runnableorchestration}), the transition with the effect {\tt enqueue(i,ORC\_CHOICE)} generates the test code {\tt oout.get(\$(ror.i)).writeObject(RunnableOrchestration.choice\_msg);} \linebreak {\tt oout.get(\$(ror.i)).flush();}, where the object {\tt oout} is a list of {\tt ObjectOutputStream}, one for each service.
Similarly, the {\tt dequeue(i)} effect generates the test code {\tt response = oin.get(\$(ror.i)).readObject();}, where {\tt oin} is a list of {\tt ObjectInputStream}.
Finally, similar to the case of testing the orchestrator, abstract values need to be instantiated in the concrete tests, and assertions are included in the tests to check whether the received messages are as expected.

For testing the interactions between two services in a distributed match, there are two further versions of the model.
In these versions, the tester is represented by the orchestrator and one of the two interactive services, either the requester or offerer in the match.

In the following, we will provide an example on how the abstract test code is generated and transformed into a concrete test.
We will focus on a specific version of the model, i.e., the one used for testing the orchestrator, and a specific set-up.

 \subsection{An example of test generation}\label{sect:example}

We now discuss an example of testing an orchestration whose set-up is dictatorial choice with centralised action. This means that the orchestrator will decide each choice and the orchestrator will act as a broker in a match between the two services. 

\paragraph{Query}
All versions of the model contains queries for testing the various configurations of the orchestrator (e.g., majoritarian choice, distributed action). We refer to~\cite{UppaalModels} for more details about these other queries. 
In the concrete tests, the orchestrator will interact with two mocked services. Furthermore, the orchestration contract automaton, a parameter of the {\tt RunnableOrchestration} object (see Figure~\ref{fig:orchestration}), is dummy and is instrumental to cover the desired portions of code of {\tt CARE}. 
In particular, the orchestration contract automaton that will be used in the concrete tests accepts the trace {\tt (!euro,-)(?coffee,!coffee)[(!euro,-)]$^*$}. 
The corresponding query used to generate the abstract test is: 

\begin{tabbing}
\hspace{1cm} \= {\tt E<>(alice.steps[0]==ORC\_CHECK \&\&} \\
\> {\tt bob.steps[1]==ORC\_CHECK \&\&} \\
\> {\tt alice.steps[2]==CENTRALISED\_OFFER \&\&}  {\it //alice offer action !euro}\\
\> {\tt alice.steps[3]==CENTRALISED\_MATCH \&\&} {\it //alice request action ?coffee}\\
\> {\tt bob.steps[4]==CENTRALISED\_OFFER \&\&} {\it //bob offer action !coffee} \\
\> {\tt alice.steps[5]==DICTATORIAL\_CHOICE \&\&} \\
\> {\tt bob.steps[6]==DICTATORIAL\_CHOICE \&\&} \\
\> {\tt alice.steps[7]==CENTRALISED\_OFFER \&\&} {\it //alice offer action !euro}\\
\> {\tt alice.steps[8]==DICTATORIAL\_CHOICE \&\&} \\
\> {\tt bob.steps[9]==DICTATORIAL\_CHOICE \&\&} \\
\> {\tt alice.steps[10]==ORC\_STOP \&\&} \\
\> {\tt bob.steps[11]==ORC\_STOP)} \\
\end{tabbing}
\noindent 
The query above is endowed with comments to pinpoint each step with the corresponding action, if any. 
An evidence to the query above simulates a run accepted by the orchestration contract automaton.
We now analyse the query.
As discussed in Section~\ref{sect:formalmodel}, the orchestration always starts by checking if the configurations
of the services and the orchestrator are matching.
Therefore, both services are instructed to check their configuration ({\tt alice.steps[0]==ORC\_CHECK} and {\tt bob.steps[1]==ORC\_CHECK}).
In this test, the configuration is of centralised action and dictatorial choice.
After the configuration checks, the orchestration starts.
In the query, Alice is instructed to execute the offer action {\tt !euro} with {\tt alice.steps[2]==CENTRALISED\_OFFER}.
This corresponds to execution of the offer label {\tt (!euro,-)} in the orchestration (Alice is the first service).
After that, the match label {\tt (?coffee,!coffee)} of the orchestration is executed.
The match starts with Alice performing the request {\tt ?coffee}, encoded in the query as
{\tt alice.steps[3]==CENTRALISED\_MATCH}.
After Alice performs the request action, it is Bob's turn to match the request of Alice with the corresponding offer action{\tt !coffee}.
In Figure~\ref{fig:RunnableOrchestratedContractSteps}, it is possible to observe that the transition of the {\tt RunnableOrchestratedContract} {\sc Uppaal} template covering the execution of an offer action in both the cases of offer label or match label (respectively, {\tt d1==NOPAYLOAD} or {\tt d1==REQUEST}, see Section~\ref{sect:formalmodel}) has in its effect the instruction {\tt steps[step]=CENTRALISED\_OFFER}.
Therefore, Bob is instructed to execute the offer action {\tt !coffee} in the query with
{\tt bob.steps[4]==CENTRALISED\_OFFER}.

After the offer label {\tt (!euro,-)} and the match label {\tt (?coffee,!coffee)} have both been executed, a choice has to be made on whether to execute the offer label {\tt (!euro,-)} or terminate (recall that the trace {\tt (!euro,-)(?coffee,!coffee)[(!euro,-)]$^*$} is accepted by the orchestration contract automaton).
In this set-up (i.e., dictatorial choice), the orchestrator performs the choice and communicates it to the services.
The orchestrator will decide the first time to execute the offer label {\tt (!euro,-)} and the second time to terminate. thus covering both cases.
The first choice is encoded in the query with {\tt alice.steps[5]==DICTATORIAL\_CHOICE} and {\tt bob.steps[6]==DICTATORIAL\_CHOICE}.
As before, the offer label is encoded in the query with {\tt alice.steps[7]==CENTRALISED\_OFFER}.
Finally, the orchestrator performs the choice to terminate.
This is encoded in the query with {\tt alice.steps[8]==DICTATORIAL\_CHOICE} and {\tt bob.steps[9]==DICTATORIAL\_CHOICE} (the orchestrator communicates the choice to the services) and {\tt alice.steps[10]==ORC\_STOP}, {\tt bob.steps[11]==ORC\_STOP} (both services terminate their execution).

\subsubsection{Test generation}
We now discuss the abstract test code generated by the query and how it is concretised into a JUnit test for the real system. 
{\sc Uppaal} also allows the insertion of abstract test code in the system declaration of a model, which will be printed out in the abstract test either at the start (i.e., pre-fix) or upon termination of the test (i.e., post-fix).
In all tests, pre-fix test code is used to declare some variables that will be utilised during the test.

\medskip
\noindent
{\it Configuration check.}
The first transitions executed during the test are those related to the configuration check.
Both Alice and Bob generate the same portion of abstract test code, reported below. 

\begin{minted}[mathescape,fontsize=\scriptsize,linenos,xleftmargin=2em]{java}
msg = (String) oin.readObject(); 

//INIT CHECK
assertEquals(msg,RunnableOrchestration.check_msg); 

String received_choiceType = (String) oin.readObject(); 
String received_actionType = (String) oin.readObject();
assertEquals(received_actionType,actionType);
assertEquals(received_choiceType,choiceType);

oout.writeObject(RunnableOrchestration.ack_msg); 
oout.flush();
//STOP CHECK
\end{minted}
\noindent 
In Figure~\ref{fig:RunnableOrchestratedContractSteps}, for both services, after the transition from the initial state {\tt Ready} is executed, the transition with the guard {\tt d1==ORC\_CHECK} is fired, followed by the subsequent transition, and finally, the transition having the effect {\tt steps[step]==ORC\_CHECK} is executed.
In order to satisfy the query these four transitions must first be executed by Alice (when {\tt step=0}) and then by Bob (when {\tt step=1}). 

The first transition performs a {\tt dequeue} operation, which appends the {\tt readObject} instruction (line~1).
The second transition executed, with the guard {\tt d1==ORC\_CHECK}, generates an assertion to check whether the message received by the orchestrator is the command for performing the configuration check (line~4).
The next transition performs two {\tt dequeue} operations, generating two {\tt readObject} instructions, followed by two assertions to check whether the two messages are those expected by the orchestrator (lines~6-9).
Recall that the orchestrator communicates the choice and action configuration (see Figure~\ref{fig:runnableorchestration}) to all services.
It is important to note here that the actual values to check ({\tt actionType} and {\tt choiceType}) will be instantiated in the concrete test, discussed next.
The fourth executed transition performs an {\tt enqueue} operation, generating the test code {\tt writeObject}, which writes the corresponding acknowledgement message into the socket (lines~11-13).

We now move to the structure of the concrete JUnit test and the concretisation of the abstract test code discussed until now. 
After that, we will discuss the remaining part of the abstract test code and its concretisation. 
The concrete test file is available at~\cite{UppaalModels} and is called { \tt DictatorialCentralisedRunnableOrchestrationTest.java}. 
All concrete tests used for testing the orchestrator follow the same schema, reported below (for the case of dictatorial choice and centralised action). 

\noindent
\captionof{listing}{The JUnit method testing the centralised dictatorial orchestrator}
\vspace{0.1em} 
\begin{minted}[mathescape,fontsize=\scriptsize,linenos,xleftmargin=2em]{java}
@Test
public void test() throws IOException, ClassNotFoundException, InterruptedException {
    Thread alice = new Thread(()->{ //mocked service
        check(8080);
        try (ServerSocketChannel server = ServerSocketChannel.open()){
            server.bind(new InetSocketAddress(8080));
            Socket socket = server.accept().socket();
            ObjectInputStream oin = new ObjectInputStream(socket.getInputStream());
            ObjectOutputStream oout = new ObjectOutputStream(socket.getOutputStream());
            oout.flush();
            uppaal_alice(oin,oout);

        } catch (IOException | ClassNotFoundException e) {
            throw new RuntimeException(e);
        }
    });

    Thread bob = new Thread(()->{ //mocked service
        check(8081);
        try (ServerSocketChannel server = ServerSocketChannel.open()){
            server.bind(new InetSocketAddress(8081));
            Socket socket = server.accept().socket();
            ObjectInputStream oin = new ObjectInputStream(socket.getInputStream());
            ObjectOutputStream oout = new ObjectOutputStream(socket.getOutputStream());
            oout.flush();
            uppaal_bob(oin,oout);
        } catch (IOException | ClassNotFoundException e) {
            throw new RuntimeException(e);
        }
    });
    alice.start();
    bob.start();
    RunnableOrchestration ror = new DictatorialChoiceRunnableOrchestration(new Agreement(),
            adc.importMSCA(dir+"Orchestration.data"),
            Arrays.asList(null,null),
            Arrays.asList(8080,8081),
            new CentralisedOrchestratorAction());

    Thread tror = new Thread(ror);
    tror.start();
    tror.join();
}
\end{minted}

The test is organised as follows.
Both Alice and Bob services are encoded as parallel threads.
Alice's thread is declared at lines~3–16, whilst Bob's thread is declared at lines~18–30.
They are both executed at lines~31–32.
After that, the orchestration is declared at lines~33–37 and executed as another parallel thread (lines~39–41).
It is important to note that, whilst Alice and Bob are two mocked services, the orchestrator is the real {\tt RunnableOrchestration} class of {\tt CARE}.
In this case, at lines~33–37, we can see that the {\tt RunnableOrchestration} object {\tt ror} is instantiated with dictatorial choice and centralised action.
The other tests for the orchestrator will have different instantiations (majoritarian choice and/or distributed action).
Furthermore, it uses the orchestration contract automaton discussed above, stored in the file {\tt Orchestration.data}. The {\tt ror} object also takes as a parameter the ports of the services.
Alice is listening at port~{\tt 8080}, and Bob is listening at port~{\tt 8081}.

Alice's thread starts by invoking the method {\tt check}, passing its port as an argument (line~4).
After that, the socket and object streams are initialised, and the method {\tt uppaal\_alice} is invoked, passing the input and output object streams as arguments. Bob's thread is equivalent.
The method {\tt check} (not reported for brevity), similarly to lines 5–10 of method {\tt test} above, opens a socket awaiting connection from the orchestrator and invokes the method {\tt uppaal\_check}. The method  {\tt uppaal\_check} contains the concretisation of the abstract test code discussed above, which is used for checking the configuration of the corresponding service.
The methods {\tt uppaal\_alice} and {\tt uppaal\_bob} (line~11 and line~26) contain the remaining concretisation of the abstract test code generated by Alice and Bob, respectively, and will be discussed in the following. 
The method {\tt uppaal\_check} is reported below. 

\noindent
\captionof{listing}{The JUnit method testing the configuration of the service}
\vspace{0.1em} 
\begin{minted}[mathescape,fontsize=\scriptsize,linenos,xleftmargin=2em]{java}
private void uppaal_check( ObjectInputStream oin, ObjectOutputStream oout) throws IOException {
    String choiceType = "Dictatorial";
    String actionType = "Centralised";
    String msg;
    msg = (String) oin.readObject();
    assertEquals(msg,RunnableOrchestration.check_msg);
    String received_choiceType = (String) oin.readObject();
    String received_actionType = (String) oin.readObject();
    assertEquals(received_actionType,actionType);
    assertEquals(received_choiceType,choiceType);
    oout.writeObject(RunnableOrchestration.ack_msg);
    oout.flush();
}
\end{minted}

It is easy to see that this method includes the abstract test case instructions discussed above.
In this case, the concretisation of the abstract test case simply instantiates the variables {\tt choiceType} and {\tt actionType} to their respective values (lines~2-3).
Note that the method {\tt uppaal\_check} will be invoked by both services.

We have discussed the main structure of the concrete test, the abstract test case instructions covering the configuration check and their concretisation. 
We now discuss the remaining instructions of the abstract test case and their concretisation in the methods  {\tt uppaal\_alice} and {\tt uppaal\_bob}.

\medskip
\noindent
{\it Offer label.}
After the orchestration check has been performed, the orchestration starts, and the offer label {\tt (!euro,-)} of the orchestration will be executed. 
The following instructions are appended to the abstract test case being generated.

\begin{minted}[mathescape,fontsize=\scriptsize,linenos,xleftmargin=2em]{java}
//centralised offerer 0

msg = (String) oin.readObject();

assertEquals(msg, expected action);

msg = (String) oin.readObject(); // reading payload centralised action
assertEquals(msg,expectedPayload);

oout.writeObject(INSERT OFFER);
oout.flush();
\end{minted}

The transition outgoing state {\tt CentralisedOffer} of the {\tt RunnableOrchestration} automaton (see Figure~\ref{fig:runnableorchestration}) has as test code  {\tt //centralised offerer \$(ror.offerer)}. 
This will append the comment at line~1 in the abstract test, indicating that the next instructions of the abstract test will cover the first offer action performed by Alice (service at index~$0$). 
This is helpful for generating the concrete test from the abstract test. 
After performing the check, both Alice and Bob are again in state {\tt Ready} (Figure~\ref{fig:RunnableOrchestratedContractSteps}). 
The first {\tt readObject} operation (line~3) is again appended by the outgoing transition from state {\tt Ready} of Alice. 
The next transition executed by Alice, with guard {\tt d1==ACTION}, will append the instruction at line~5. 
The assertion will check whether the message received by the orchestrator will be the expected action, which will be concretised in the concrete test with the value {\tt "euro"} (i.e., the orchestrator is invoking the method {\tt euro} of Alice). 

After that, Alice automaton will execute the centralised action, and the subsequent transition will perform a {\tt dequeue} operation, appending to the abstract test case the instructions at lines~7-8. 
In the assertion, {\tt expectedPayload} is a placeholder that will be concretised with value {\tt null}. 
Recall that in {\tt CARE} each action of the contract automaton of each service is in correspondence with a method of the corresponding {\tt service} class (see Section~\ref{sect:background}), which has input parameters and a returned value. 
In this case, the parameters and returned values of the method in correspondence with the offer {\tt euro} of Alice are dummy (i.e., concretised with {\tt null} values). 
The next transition with effect  {\tt enqueue}  is executed by Alice. 
This transition also has in the effect the instruction {\tt steps[step]=CENTRALISED\_OFFER}, thus fulfilling the query (at this point of the simulation it holds {\tt step=2}). 
The execution of the transition will append instructions at lines~10-11 to the abstract test case. 
Indeed, the  {\tt enqueue} operation corresponds to a {\tt writeObject} operation in the test. 
The placeholder {\tt INSERT OFFER} will be replaced by the concrete value {\tt null} in the concrete test. 

This concludes the test code for the first offer label {\tt (!euro,-)} of the orchestration.

\medskip
\noindent
{\it Match label.} 
Next, the match label {\tt (?coffee,!coffee)} is executed. 
The instructions appended to the abstract test case during the execution of the match label are reported below.

\begin{minted}[mathescape,fontsize=\scriptsize,linenos,xleftmargin=2em]{java}
//requester: 0

msg = (String) oin.readObject();

assertEquals(msg, expected action);

msg = (String) oin.readObject(); // reading payload centralised action
assertEquals(msg,expectedPayload);

oout.writeObject(INSERT REQUEST);
oout.flush();

//matching offerer: 1

msg = (String) oin.readObject();

assertEquals(msg, expected action);

msg = (String) oin.readObject(); // reading payload centralised action
assertEquals(msg,expectedPayload);

oout.writeObject(INSERT OFFER);
oout.flush();

//requester: 0

msg = (String) oin.readObject(); //receiving the matching offer
assertEquals(msg, expected payload);
\end{minted} 

The three comments (lines~1,13,25) are appended by the {\tt RunnableOrchestration} automaton, after state {\tt CentralisedMatch} is reached (see Figure~\ref{fig:runnableorchestration}), and are useful to identify the instructions generated by Alice ({\tt requester 0}) and Bob ({\tt matching offerer~1}). 
The instructions at lines~3-11 are appended by Alice when executing the request action {\tt ?coffee}. 
The placeholder {\tt expectedPayload} at line~8 will be concretised with the value {\tt "coffee"} (i.e., the orchestrator is invoking the method {\tt coffee} of Alice), whilst the placeholder at line~10 will be replaced with the concrete value {\tt "request payload"} (i.e., the request sent by Alice to Bob is the dummy String  {\tt "request payload"}). 
Indeed, although the services are mocked by the test, in case of a match these values will be handled by the real {\tt CARE} orchestrator, which will act as a broker between the two services. 
After executing the transition with effect {\tt steps[step]=CENTRALISED\_MATCH} (Figure~\ref{fig:RunnableOrchestratedContractSteps}), thus fulfilling the query, Alice waits for Bob to send the offer. 
The instructions at lines~15-23 are appended by Bob when executing the offer action {\tt !coffee}. 
Note that these are identical to the previous offer action {\tt !euro} fired by Alice, because the same transitions of Figure~\ref{fig:RunnableOrchestratedContractSteps} have been executed. 
The placeholders at line~17, line~20 and line~22 will be replaced by the concrete values {\tt "coffee"} (the orchestrator invokes the method {\tt coffee} of Bob), {\tt "request payload"} (the payload sent by Alice to Bob is checked) and {\tt "offer payload"} (the dummy reply of Bob to Alice), respectively.  
After that, the instructions at lines~27-28 are executed by Alice for completing the request action. 
The placeholder at line~28 will be replaced by the concrete value {\tt "offer payload"} (the reply of Bob to Alice is checked). 

This concludes the match  {\tt (?coffee,!coffee)}. 

\medskip
\noindent
{\it Choice.} 
As stated earlier, the orchestrator will now perform a choice on whether to continue and execute the offer label
{\tt (!euro,-)} or terminate. 
For brevity, we only report the abstract test code dumped by a service when executing a dictatorial choice, reported below.

\begin{minted}[mathescape,fontsize=\scriptsize,linenos,xleftmargin=2em]{java}
msg = (String) oin.readObject();

assertEquals(msg,RunnableOrchestration.choice_msg);
\end{minted}

The instruction at line~1 is again appended when executing the {\tt dequeue} effect outgoing state {\tt Ready} (Figure~\ref{fig:RunnableOrchestratedContractSteps}). 
After that, the transition with guard {\tt d1 == ORC\_CHOICE} is executed, which will append the test code at line~3. 
The next executed transition, with effect {\tt steps[step]=DICTATORIAL\_CHOICE}, has no test code generated but it is used to fulfill the query. 
No other test code is generated for the dictatorial choice, as the orchestrator decides on its own the next move. 
This test code is generated by both Alice and Bob, performing in turn the choice. 
In this example, after the first dictatorial choice, the orchestrator will execute again the offer label {\tt (!euro,-)}, and again a dictatorial choice. 
We have already covered the test code that will be generated. 

\medskip
\noindent
{\it Termination.}
After that, the orchestrator decides to terminate. 
The test code generated by each service is:  

\begin{minted}[mathescape,fontsize=\scriptsize,linenos,xleftmargin=2em]{java}
msg = (String) oin.readObject();

assertEquals(msg,RunnableOrchestration.stop_msg);
\end{minted}

The instruction at line~1 is the first {\tt dequeue} operation as above. 
The instruction at line~3 is instead appended by the test code of the transition with guard {\tt d1==ORC\_STOP} (Figure~\ref{fig:RunnableOrchestratedContractSteps}). This transition has also in its effect the instruction 
{\tt steps[step]=ORC\_STOP}, thus fulfilling the query. 
Note that both values {\tt RunnableOrchestration.choice\_msg} and {\tt RunnableOrchestration.stop\_msg} are constants 
of the class {\tt RunnableOrchestration} of {\tt CARE}, used to identify the two commands of the orchestrator. 

We have already discussed how the values of the abstract test code are concretised. 
For completeness, we report below the code of the method {\tt uppaal\_alice} discussed before, which contains  
test code for the service Alice, obtained by concretising the abstract test code as discussed above.

\noindent
\captionof{listing}{The JUnit method testing the Alice service}
\vspace{0.1em}
\begin{minted}[mathescape,fontsize=\scriptsize,linenos,xleftmargin=2em]{java}
private void uppaal_alice( ObjectInputStream oin, ObjectOutputStream oout) throws IOException {
    String msg;
    msg = (String) oin.readObject();
    assertEquals(msg, "euro");
    msg = (String) oin.readObject(); // reading payload centralised action
    assertEquals(msg,null);
    oout.writeObject(null);
    oout.flush();
    
    msg = (String) oin.readObject();
    assertEquals(msg, "coffee");
    msg = (String) oin.readObject(); // reading payload centralised action
    assertEquals(msg,null);
    oout.writeObject("request payload");
    oout.flush();

    msg = (String) oin.readObject(); //receiving the matching offer
    assertEquals(msg, "offer payload");

    msg = (String) oin.readObject();
    assertEquals(msg,RunnableOrchestration.choice_msg);

    msg = (String) oin.readObject();

    //if the orchestrator decides not to terminate
    while (msg.equals("euro")) {
        msg = (String) oin.readObject(); // reading payload centralised action
        assertEquals(msg, null);
        oout.writeObject(null);
        oout.flush();
        msg = (String) oin.readObject();
        assertEquals(msg, RunnableOrchestration.choice_msg);
        msg = (String) oin.readObject();
    }

    assertEquals(msg,RunnableOrchestration.stop_msg);
}
\end{minted}

Note that the above code allows Alice to perform more than a single offer {\tt "euro"} before terminating (lines 26~34). 
In this way it is possible to handle all possible traces of the orchestration automaton, and all internal choices of the orchestrator. 

Similarly, we report below the code of the method {\tt uppaal\_bob}.

\noindent
\captionof{listing}{The JUnit method testing the Bob service}
\vspace{0.1em}
\begin{minted}[mathescape,fontsize=\scriptsize,linenos,xleftmargin=2em]{java}
private void uppaal_bob(ObjectInputStream oin, ObjectOutputStream oout) throws IOException {
    String msg;

    msg = (String) oin.readObject();
    assertEquals(msg, "coffee");
    msg = (String) oin.readObject(); // reading payload centralised action
    assertEquals(msg,"request payload");
    oout.writeObject("offer payload");
    oout.flush();

    msg = (String) oin.readObject();
    assertEquals(msg,RunnableOrchestration.choice_msg);

    //in case the orchestrator decides not to terminate
    do {
        msg = (String) oin.readObject();
    } while (msg.equals(RunnableOrchestration.choice_msg));

    assertEquals(msg,RunnableOrchestration.stop_msg);

}
\end{minted}

Similar to Alice, also the code of Bob allows to handle many unfoldings of the final loop of the orchestration. 

This concludes the example of test generation.
Recall that this is one query of one version of the model (for testing the orchestrator).
There are other queries used for testing the other configurations (e.g., majoritarian choice, distributed offer) and other versions of the model for testing the other components. 
However, the generation of abstract tests and their concretisation follows the same approach covered by this example. 
All these versions and the concrete tests are available at~\cite{UppaalModels}.



\subsection{Coverage}
\begin{figure}[t]
\includegraphics[width=0.8\columnwidth]{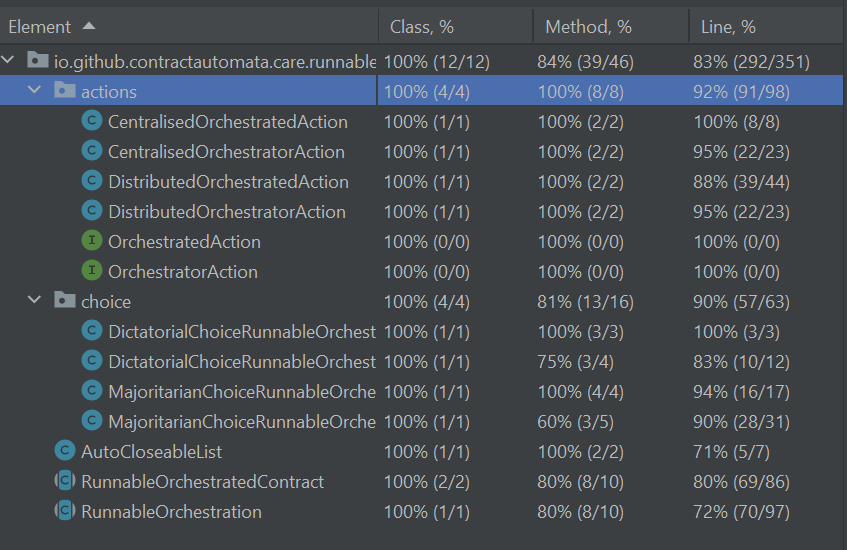}
\caption{The coverage information generated by IntelliJ}
\label{fig:coverage}
\end{figure}

The coverage information generated by IntelliJ is in Figure~\ref{fig:coverage}. 
 The code coverage indicates that the concrete tests derived from the model cover a significant portion of the source code. 
 This suggests that the model is not excessively abstract compared to the actual implementation. 
Furthermore, from the queries we know that the generated abstract tests cover all transitions of the model and all interactions between the orchestrator and the services. 

On a side note, the modelling and verification allowed us to 
 fix some issues (see Section~\ref{sect:analysis}). 
This would not have been possible in case the abstraction were too coarse.


\section{Conclusion and Future Work}\label{sect:conclusion}
\vspace{-0.08cm}
We have presented the formal modelling and verification of the Contract Automata Runtime Environment ({\tt CARE}), an open-source platform.
%
Both statistical and exhaustive model checking techniques have played a crucial role in formally verifying numerous desired properties of the modeled system, such as the absence of deadlocks, while also enhancing the accuracy of the formal model. Statistical model checking has also been employed to fine-tune parameter settings within the formal model, such as the buffer size. 
The adequacy of the abstract model has been described, and the transitions of the formal model have been linked to the corresponding lines of source code. The tests generated from the formal model have been employed to test the source code. 

At present, the different artifacts such as the model, source code, tests, and tracing information are manually kept aligned. This process demands substantial effort whenever a new version of {\tt CARE} is introduced, as each artifact needs to be updated accordingly. Future work involves studying techniques for automatic alignment of these artifacts. 
Furthermore, it would be interesting to exploit the facilities provided by the recently introduced tool {\tt Uppex}~\cite{DBLP:journals/corr/abs-2310-20395} to factorize the different variants of the model discussed in the paper under a single, configurable {\tt Uppex} model, to automate the selection of a configuration and more generally to facilitate the collaboration between the software developer and the {\sc Uppaal} modeller, whenever they are different persons.

\paragraph{\bf Acknowledgment}
Part of this study was carried out within the 
MUR PRIN 2022 PNRR P2022A492B project ADVENTURE (ADVancEd iNtegraTed evalUation of Railway systEms) funded by the European Union - NextGenerationEU, and the CNR project ``Formal Methods in Software Engineering 2.0'', CUP B53C24000720005.

\bibliographystyle{alphaurl}
\bibliography{bib}

\end{document}

%% file: macros.tex
\usepackage{amsmath,amssymb,amstext} 
\usepackage[misc]{ifsym}
\usepackage{xcolor}
\usepackage{graphicx}
\usepackage{wrapfig}

\usepackage{doi,url}

\usepackage{xifthen}
\newcommand{\inv}[1][]{%
\ifthenelse{\equal{#1}{}}{I}{I(#1)}%
}
\newcommand{\con}[1][]{%
\ifthenelse{\equal{#1}{}}{\chi}{\chi(#1)}%
}

\renewcommand{\epsilon}{\varepsilon}

\newcommand{\hbra}{
\hbox to 1 \textwidth{\vrule width0.3mm height 1.8mm depth-0.3mm
                    \leaders\hrule height1.8mm depth-1.5mm\hfill
                    \vrule width0.3mm height 1.8mm depth-0.3mm}}
\newcommand{\hket}{
\hbox to 1 \textwidth{\vrule width0.3mm height1.5mm
                    \leaders\hrule height0.3mm\hfill
                    \vrule width0.3mm height1.5mm}}

\makeatletter
\def\thickhline{%
  \noalign{\ifnum0=`}\fi\hrule \@height \thickarrayrulewidth \futurelet
   \reserved@a\@xthickhline}
\def\@xthickhline{\ifx\reserved@a\thickhline
               \vskip\doublerulesep
               \vskip-\thickarrayrulewidth
             \fi
      \ifnum0=`{\fi}}
\makeatother
\newlength{\thickarrayrulewidth}
\usepackage[most]{tcolorbox}
\newtcolorbox{markbox}{
  enhanced,
  breakable,
  size=minimal,
  parbox=false,
  after={\par},
  before upper={\indent},
  colback=white,
  overlay = {
	 \draw[line width=2pt]
	 (frame.north east) -| ([xshift=3mm]frame.east) |-(frame.south east);
  },
  overlay first={\draw[line width=2pt] (frame.north east) -| ([xshift=3mm]frame.south east);},
  overlay middle={\draw[line width=2pt] ([xshift=3mm]frame.north east) -- ([xshift=3mm]frame.south east);},
  overlay last={\draw[line width=2pt] ([xshift=3mm]frame.north east) |- (frame.south east);},
}

\usepackage{wrapfig}

\usepackage{lipsum}  

%% file: main.bbl
\newcommand{\etalchar}[1]{$^{#1}$}
\begin{thebibliography}{WVHFM06}

\bibitem[ABB{\etalchar{+}}16]{DBLP:series/lncs/10001}
Wolfgang Ahrendt, Bernhard Beckert, Richard Bubel, Reiner H{\"{a}}hnle,
  Peter~H. Schmitt, and Mattias Ulbrich, editors.
\newblock {\em Deductive Software Verification -- The KeY Book: From Theory to
  Practice}, volume 10001 of {\em LNCS}.
\newblock Springer, 2016.
\newblock \href {https://doi.org/10.1007/978-3-319-49812-6}
  {\path{doi:10.1007/978-3-319-49812-6}}.

\bibitem[ABZ13]{ABZ13}
L.~Acciai, M.~Boreale, and G.~Zavattaro.
\newblock Behavioural contracts with request-response operations.
\newblock {\em Sci. Comp. Program.}, 78(2):248--267, 2013.
\newblock \href {https://doi.org/10.1016/j.scico.2011.10.007}
  {\path{doi:10.1016/j.scico.2011.10.007}}.

\bibitem[AD94]{AD94}
R.~Alur and D.~Dill.
\newblock {A Theory of Timed Automata}.
\newblock {\em Theoret. Comp. Sci.}, 126(2):183--235, 1994.
\newblock \href {https://doi.org/10.1016/0304-3975(94)90010-8}
  {\path{doi:10.1016/0304-3975(94)90010-8}}.

\bibitem[AD24]{DBLP:conf/coordination/AlsubhiD24}
Arwa~Hameed Alsubhi and Ornela Dardha.
\newblock Coconut: Typestates for embedded systems.
\newblock In Ilaria Castellani and Francesco Tiezzi, editors, {\em Coordination
  Models and Languages - 26th {IFIP} {WG} 6.1 International Conference,
  {COORDINATION} 2024, Held as Part of the 19th International Federated
  Conference on Distributed Computing Techniques, DisCoTec 2024, Groningen, The
  Netherlands, June 17-21, 2024, Proceedings}, volume 14676 of {\em Lecture
  Notes in Computer Science}, pages 219--238. Springer, 2024.
\newblock \href {https://doi.org/10.1007/978-3-031-62697-5_12}
  {\path{doi:10.1007/978-3-031-62697-5_12}}.

\bibitem[Basa]{CATAPPurl}
Davide Basile.
\newblock {CATApp}.
\newblock URL:
  \url{https://github.com/contractautomataproject/ContractAutomataApp}.

\bibitem[Basb]{UppaalModels}
Davide Basile.
\newblock Formal analysis of the {C}ontract {A}utomata {R}untime {E}nvironment
  with {U}ppaal, complementary material.
\newblock \href {https://doi.org/10.5281/zenodo.14671729}
  {\path{doi:10.5281/zenodo.14671729}}.

\bibitem[Bas24]{DBLP:conf/coordination/Basile24}
Davide Basile.
\newblock Modelling, verifying and testing the contract automata runtime
  environment with uppaal.
\newblock In Ilaria Castellani and Francesco Tiezzi, editors, {\em Coordination
  Models and Languages - 26th {IFIP} {WG} 6.1 International Conference,
  {COORDINATION} 2024, Held as Part of the 19th International Federated
  Conference on Distributed Computing Techniques, DisCoTec 2024, Groningen, The
  Netherlands, June 17-21, 2024, Proceedings}, volume 14676 of {\em Lecture
  Notes in Computer Science}, pages 93--110. Springer, 2024.
\newblock \href {https://doi.org/10.1007/978-3-031-62697-5_6}
  {\path{doi:10.1007/978-3-031-62697-5_6}}.

\bibitem[BBG{\etalchar{+}}22]{BACCHIANI2022102844}
Lorenzo Bacchiani, Mario Bravetti, Marco Giunti, Jo{\~{a}}o Mota, and
  Ant{\'{o}}nio Ravara.
\newblock A {Java} typestate checker supporting inheritance.
\newblock {\em Sci. Comput. Program.}, 221, 2022.
\newblock \href {https://doi.org/10.1016/j.scico.2022.102844}
  {\path{doi:10.1016/j.scico.2022.102844}}.

\bibitem[BCZ15]{BCZ15}
Massimo Bartoletti, Tiziana Cimoli, and Roberto Zunino.
\newblock {Compliance in Behavioural Contracts: A Brief Survey}.
\newblock In {\em Programming Languages with Applications to Biology and
  Security}, volume 9465 of {\em LNCS}, pages 103--121. Springer, 2015.
\newblock \href {https://doi.org/10.1007/978-3-319-25527-9_9}
  {\path{doi:10.1007/978-3-319-25527-9_9}}.

\bibitem[BDF16]{BDF16}
Davide Basile, Pierpaolo Degano, and Gian-Luigi Ferrari.
\newblock Automata for specifying and orchestrating service contracts.
\newblock {\em Log. Methods Comput. Sci.}, 12(4):6:1--6:51, 2016.
\newblock \href {https://doi.org/10.2168/LMCS-12(4:6)2016}
  {\path{doi:10.2168/LMCS-12(4:6)2016}}.

\bibitem[BDL{\etalchar{+}}06]{BDLHPYH06}
G.~Behrmann, A.~David, K.~G. Larsen, J.~H{\aa}kansson, P.~Pettersson, W.~Yi,
  and M.~Hendriks.
\newblock {UPPAAL}~4.0.
\newblock In {\em Proceedings 3rd International Conference on the Quantitative
  Evaluation of SysTems (QEST)}, pages 125--126. IEEE, 2006.
\newblock \href {https://doi.org/10.1109/QEST.2006.59}
  {\path{doi:10.1109/QEST.2006.59}}.

\bibitem[BLMT08]{BLMT08}
R.~Bruni, I.~Lanese, H.~C. Melgratti, and E.~Tuosto.
\newblock {Multiparty Sessions in {SOC}}.
\newblock In {\em COORDINATION}, volume 5052 of {\em LNCS}, pages 67--82.
  Springer, 2008.
\newblock \href {https://doi.org/10.1007/978-3-540-68265-3_5}
  {\path{doi:10.1007/978-3-540-68265-3_5}}.

\bibitem[BMF23]{fmics/BasileMF23}
Davide Basile, Franco Mazzanti, and Alessio Ferrari.
\newblock Experimenting with formal verification and model-based development in
  railways: The case of {UMC} and sparx enterprise architect.
\newblock In Alessandro Cimatti and Laura Titolo, editors, {\em Formal Methods
  for Industrial Critical Systems (FMICS)}, volume 14290 of {\em LNCS}, pages
  1--21. Springer, 2023.
\newblock \href {https://doi.org/10.1007/978-3-031-43681-9_1}
  {\path{doi:10.1007/978-3-031-43681-9_1}}.

\bibitem[Bou15]{Boulanger}
Jean-Louis Boulanger.
\newblock {Tool Qualification}.
\newblock In {\em CENELEC 50128 and IEC 62279 Standards}, chapter~9, pages
  287--308. John Wiley \& Sons, 2015.
\newblock \href {https://doi.org/10.1002/9781119005056.ch9}
  {\path{doi:10.1002/9781119005056.ch9}}.

\bibitem[BtB21]{BasileB21}
Davide Basile and Maurice~H. ter Beek.
\newblock {A Clean and Efficient Implementation of Choreography Synthesis for
  Behavioural Contracts}.
\newblock In Ferruccio Damiani and Ornela Dardha, editors, {\em COORDINATION},
  volume 12717 of {\em LNCS}, pages 225--238. Springer, 2021.
\newblock \href {https://doi.org/10.1007/978-3-030-78142-2_14}
  {\path{doi:10.1007/978-3-030-78142-2_14}}.

\bibitem[BtB22]{BB22OSP}
Davide Basile and Maurice~H. ter Beek.
\newblock {Contract Automata Library}.
\newblock {\em Sci. Comput. Program.}, 221:102841, 2022.
\newblock URL:
  \url{https://github.com/contractautomataproject/ContractAutomataLib}, \href
  {https://doi.org/10.1016/j.scico.2022.102841}
  {\path{doi:10.1016/j.scico.2022.102841}}.

\bibitem[BtB23a]{DBLP:conf/fm/BasileB23}
Davide Basile and Maurice~H. ter Beek.
\newblock {A Runtime Environment for Contract Automata}.
\newblock In Marsha Chechik, Joost{-}Pieter Katoen, and Martin Leucker,
  editors, {\em FM}, volume 14000 of {\em LNCS}, pages 550--567. Springer,
  2023.
\newblock URL: \url{https://github.com/contractautomataproject/CARE}, \href
  {https://doi.org/10.1007/978-3-031-27481-7_31}
  {\path{doi:10.1007/978-3-031-27481-7_31}}.

\bibitem[BtB23b]{DBLP:journals/corr/abs-2308-10651}
Davide Basile and Maurice~H. ter Beek.
\newblock Research challenges in orchestration synthesis.
\newblock In Cl{\'{e}}ment Aubert, Cinzia~Di Giusto, Simon Fowler, and Larisa
  Safina, editors, {\em Proceedings of the 16th Interaction and Concurrency
  Experience (ICE)}, volume 383 of {\em EPTCS}, pages 73--90, 2023.
\newblock \href {https://doi.org/10.4204/EPTCS.383.5}
  {\path{doi:10.4204/EPTCS.383.5}}.

\bibitem[BtB24]{DBLP:journals/jlap/BasileB24}
Davide Basile and Maurice~H. ter Beek.
\newblock Advancing orchestration synthesis for contract automata.
\newblock {\em J. Log. Algebraic Methods Program.}, 141:100998, 2024.
\newblock \href {https://doi.org/10.1016/J.JLAMP.2024.100998}
  {\path{doi:10.1016/J.JLAMP.2024.100998}}.

\bibitem[BtBC18]{BBC18}
Davide Basile, Maurice~H. ter Beek, and Vincenzo Ciancia.
\newblock {Statistical Model Checking of a Moving Block Railway Signalling
  Scenario with {\sc Uppaal} {SMC}}.
\newblock In Tiziana Margaria and Bernhard Steffen, editors, {\em ISoLA},
  volume 11245 of {\em LNCS}, pages 372--391. Springer, 2018.
\newblock \href {https://doi.org/10.1007/978-3-030-03421-4_24}
  {\path{doi:10.1007/978-3-030-03421-4_24}}.

\bibitem[BtBD{\etalchar{+}}20]{BBDLFGD20}
Davide Basile, Maurice~H. ter Beek, Pierpaolo Degano, Axel Legay, Gian-Luigi
  Ferrari, Stefania Gnesi, and Felicita {Di Giandomenico}.
\newblock {Controller synthesis of service contracts with variability}.
\newblock {\em Sci. Comput. Program.}, 187:102344, 2020.
\newblock \href {https://doi.org/10.1016/j.scico.2019.102344}
  {\path{doi:10.1016/j.scico.2019.102344}}.

\bibitem[BtBFL22]{DBLP:journals/sttt/BasileBFL22}
Davide Basile, Maurice~H. ter Beek, Alessio Ferrari, and Axel Legay.
\newblock {Exploring the {ERTMS/ETCS} full moving block specification: an
  experience with formal methods}.
\newblock {\em Int. J. Softw. Tools Technol. Transf.}, 24(3):351--370, 2022.
\newblock \href {https://doi.org/10.1007/s10009-022-00653-3}
  {\path{doi:10.1007/s10009-022-00653-3}}.

\bibitem[BtBL20]{BBL20}
Davide Basile, Maurice~H. ter Beek, and Axel Legay.
\newblock {Strategy Synthesis for Autonomous Driving in a Moving Block Railway
  System with \textsc{Uppaal Stratego}}.
\newblock In Alexey Gotsman and Ana Sokolova, editors, {\em FORTE}, volume
  12136 of {\em LNCS}, pages 3--21. Springer, 2020.
\newblock \href {https://doi.org/10.1007/978-3-030-50086-3_1}
  {\path{doi:10.1007/978-3-030-50086-3_1}}.

\bibitem[BtBP20]{BBP20}
Davide Basile, Maurice~H. ter Beek, and Rosario Pugliese.
\newblock Synthesis of orchestrations and choreographies: {B}ridging the gap
  between supervisory control and coordination of services.
\newblock {\em Log. Methods Comput. Sci.}, 16(2):9:1--9:29, 2020.
\newblock \href {https://doi.org/10.23638/LMCS-16(2:9)2020}
  {\path{doi:10.23638/LMCS-16(2:9)2020}}.

\bibitem[CDCP12]{CDP12}
G.~Castagna, M.~Dezani-Ciancaglini, and L.~Padovani.
\newblock {On Global Types and Multi-Party Sessions}.
\newblock {\em Log. Methods Comput. Sci.}, 8(1:24):1--45, 2012.
\newblock \href {https://doi.org/10.2168/LMCS-8(1:24)2012}
  {\path{doi:10.2168/LMCS-8(1:24)2012}}.

\bibitem[CGMZ21]{DBLP:conf/secdev/CluzelGMZ21}
Guillaume Cluzel, Kyriakos Georgiou, Yannick Moy, and Cl{\'{e}}ment Zeller.
\newblock Layered formal verification of a {TCP} stack.
\newblock In {\em SecDev}, pages 86--93. {IEEE}, 2021.
\newblock \href {https://doi.org/10.1109/SecDev51306.2021.00028}
  {\path{doi:10.1109/SecDev51306.2021.00028}}.

\bibitem[CGP09]{CGP09}
G.~Castagna, N.~Gesbert, and L.~Padovani.
\newblock {A Theory of Contracts for Web Services}.
\newblock {\em ACM Trans. Program. Lang. Syst.}, 31(5):19:1--19:61, 2009.
\newblock \href {https://doi.org/10.1145/1538917.1538920}
  {\path{doi:10.1145/1538917.1538920}}.

\bibitem[CHLR23]{10.1145/3595376}
H\'{e}l\'{e}ne Coullon, Ludovic Henrio, Fr\'{e}d\'{e}ric Loulergue, and Simon
  Robillard.
\newblock Component-based distributed software reconfiguration:a
  verification-oriented survey.
\newblock {\em ACM Comput. Surv.}, 56(1), August 2023.
\newblock \href {https://doi.org/10.1145/3595376} {\path{doi:10.1145/3595376}}.

\bibitem[dAH01]{AH01}
L.~de~Alfaro and T.~Henzinger.
\newblock {Interface Automata}.
\newblock In {\em ESEC/FSE}, pages 109--120. ACM, 2001.
\newblock \href {https://doi.org/10.1145/503209.503226}
  {\path{doi:10.1145/503209.503226}}.

\bibitem[DCd10]{DD09}
M.~Dezani-Ciancaglini and U.~de'Liguoro.
\newblock {Sessions and Session Types: An Overview}.
\newblock In {\em WS-FM}, volume 6194 of {\em LNCS}, pages 1--28. Springer,
  2010.
\newblock \href {https://doi.org/10.1007/978-3-642-14458-5_1}
  {\path{doi:10.1007/978-3-642-14458-5_1}}.

\bibitem[dlCGMS09]{DBLP:journals/sttt/CamaraGMS09}
Pedro de~la C{\'{a}}mara, Mar{\'{\i}}a{-}del{-}Mar Gallardo, Pedro Merino, and
  David San{\'{a}}n.
\newblock Checking the reliability of socket based communication software.
\newblock {\em Int. J. Softw. Tools Technol. Transf.}, 11(5):359--374, 2009.
\newblock \href {https://doi.org/10.1007/s10009-009-0112-7}
  {\path{doi:10.1007/s10009-009-0112-7}}.

\bibitem[DLL{\etalchar{+}}10]{DLLNW10}
A.~David, K.~G. Larsen, A.~Legay, U.~Nyman, and A.~Wasowski.
\newblock {Timed I/O Automata: A Complete Specification Theory for Real-time
  Systems}.
\newblock In {\em HSCC}, pages 91--100. ACM, 2010.
\newblock \href {https://doi.org/10.1145/1755952.1755967}
  {\path{doi:10.1145/1755952.1755967}}.

\bibitem[DLL{\etalchar{+}}15]{DLLMP15}
A.~David, K.~G. Larsen, A.~Legay, M.~Miku{\v{c}}ionis, and D.~B. Poulsen.
\newblock {\textsc{Uppaal} SMC tutorial}.
\newblock {\em Int. J. Softw. Tools Technol. Transf.}, 17(4):397--415, 2015.
\newblock \href {https://doi.org/10.1007/s10009-014-0361-y}
  {\path{doi:10.1007/s10009-014-0361-y}}.

\bibitem[FtB22]{10.1145/3520480}
Alessio Ferrari and Maurice~H. ter Beek.
\newblock {Formal Methods in Railways: {A} Systematic Mapping Study}.
\newblock {\em ACM Comput. Surv.}, 2022.
\newblock \href {https://doi.org/10.1145/3520480} {\path{doi:10.1145/3520480}}.

\bibitem[FZL18]{DBLP:conf/tase/FeiZL18}
Yuan Fei, Huibiao Zhu, and Xin Li.
\newblock Modeling and verification of {NLSR} protocol using {UPPAAL}.
\newblock In Jun Pang, Chenyi Zhang, Jifeng He, and Jian Weng, editors, {\em
  TASE}, pages 108--115. {IEEE} Computer Society, 2018.
\newblock \href {https://doi.org/10.1109/TASE.2018.00022}
  {\path{doi:10.1109/TASE.2018.00022}}.

\bibitem[GJP{\etalchar{+}}22]{DBLP:journals/sttt/GuJPSEL22}
Rong Gu, Peter~Gj{\o}l Jensen, Danny~B{\o}gsted Poulsen, Cristina Seceleanu,
  Eduard Enoiu, and Kristina Lundqvist.
\newblock Verifiable strategy synthesis for multiple autonomous agents: a
  scalable approach.
\newblock {\em Int. J. Softw. Tools Technol. Transf.}, 24(3):395--414, 2022.
\newblock \href {https://doi.org/10.1007/s10009-022-00657-z}
  {\path{doi:10.1007/s10009-022-00657-z}}.

\bibitem[GR17]{Gay20171}
Simon Gay and António Ravara, editors.
\newblock {\em Behavioural Types: from Theory to Tools}.
\newblock River, 2017.
\newblock \href {https://doi.org/10.13052/rp-9788793519817}
  {\path{doi:10.13052/rp-9788793519817}}.

\bibitem[GtBvdP20]{GBP20}
Hubert Garavel, Maurice~H. ter Beek, and Jaco van~de Pol.
\newblock {The 2020 Expert Survey on Formal Methods}.
\newblock In M.H. ter Beek and D.~Ni{\v{c}}kovi{\'{c}}, editors, {\em FMICS},
  volume 12327 of {\em LNCS}, pages 3--69. Springer, 2020.
\newblock \href {https://doi.org/10.1007/978-3-030-58298-2_1}
  {\path{doi:10.1007/978-3-030-58298-2_1}}.

\bibitem[HL23]{DBLP:conf/medi/HammamiL23}
Mohamed~Amin Hammami and Mariam Lahami.
\newblock Model-based testing approach for {EIP-1559} ethereum smart contracts.
\newblock In Mohamed Mosbah, M.~Tahar Kechadi, Ladjel Bellatreche, and
  Fa{\"{\i}}ez Gargouri, editors, {\em Model and Data Engineering - 12th
  International Conference, {MEDI} 2023, Sousse, Tunisia, November 2-4, 2023,
  Proceedings}, volume 14396 of {\em Lecture Notes in Computer Science}, pages
  44--57. Springer, 2023.
\newblock \href {https://doi.org/10.1007/978-3-031-49333-1_4}
  {\path{doi:10.1007/978-3-031-49333-1_4}}.

\bibitem[HLM{\etalchar{+}}08]{DBLP:conf/fortest/HesselLMNPS08}
Anders Hessel, Kim~Guldstrand Larsen, Marius Mikucionis, Brian Nielsen, Paul
  Pettersson, and Arne Skou.
\newblock Testing real-time systems using {UPPAAL}.
\newblock In Robert~M. Hierons, Jonathan~P. Bowen, and Mark Harman, editors,
  {\em Formal Methods and Testing, An Outcome of the {FORTEST} Network, Revised
  Selected Papers}, volume 4949 of {\em Lecture Notes in Computer Science},
  pages 77--117. Springer, 2008.
\newblock \href {https://doi.org/10.1007/978-3-540-78917-8_3}
  {\path{doi:10.1007/978-3-540-78917-8_3}}.

\bibitem[HLV{\etalchar{+}}16]{Hut16}
H.~H\"{u}ttel, I.~Lanese, V.~T. Vasconcelos, L.~Caires, M.~Carbone, P.~{-}~M.
  Deni{\'e}lou, D.~Mostrous, L.~Padovani, A.~Ravara, E.~Tuosto, H.~{Torres
  Vieira}, and G.~Zavattaro.
\newblock {Foundations of Session Types and Behavioural Contracts}.
\newblock {\em ACM Comput. Surv.}, 49(1):3:1--3:36, 2016.
\newblock \href {https://doi.org/10.1145/2873052} {\path{doi:10.1145/2873052}}.

\bibitem[HMG{\etalchar{+}}23]{sys.21679}
Benedek Horváth, Vince Molnár, Bence Graics, \'Akos Hajdu, István Ráth,
  \'Akos Horváth, Robert Karban, Gelys Trancho, and Zoltán Micskei.
\newblock Pragmatic verification and validation of industrial executable sysml
  models.
\newblock {\em Systems Engineering}, 2023.
\newblock \href {https://doi.org/10.1002/sys.21679}
  {\path{doi:10.1002/sys.21679}}.

\bibitem[HYC08]{HYC08}
K.~Honda, N.~Yoshida, and M.~Carbone.
\newblock {Multiparty Asynchronous Session Types}.
\newblock In {\em POPL}, pages 273--284. ACM, 2008.
\newblock \href {https://doi.org/10.1145/1328438.1328472}
  {\path{doi:10.1145/1328438.1328472}}.

\bibitem[Jav]{JavaSocket}
URL: \url{https://docs.oracle.com/javase/7/docs/api/java/net/Socket.html}.

\bibitem[KDPG18]{DBLP:journals/scp/KouzapasDPG18}
Dimitrios Kouzapas, Ornela Dardha, Roly Perera, and Simon~J. Gay.
\newblock Typechecking protocols with {Mungo} and {StMungo}: {A} session type
  toolchain for {Java}.
\newblock {\em Sci. Comput. Program.}, 155:52--75, 2018.
\newblock \href {https://doi.org/10.1016/j.scico.2017.10.006}
  {\path{doi:10.1016/j.scico.2017.10.006}}.

\bibitem[KLN{\etalchar{+}}15]{DBLP:conf/fmics/KimLNMO15}
Jin~Hyun Kim, Kim~G. Larsen, Brian Nielsen, Marius Mikucionis, and Petur Olsen.
\newblock Formal analysis and testing of real-time automotive systems using
  {UPPAAL} tools.
\newblock In Manuel N{\'{u}}{\~{n}}ez and Matthias G{\"{u}}demann, editors,
  {\em Formal Methods for Industrial Critical Systems - 20th International
  Workshop, {FMICS} 2015, Oslo, Norway, June 22-23, 2015 Proceedings}, volume
  9128 of {\em Lecture Notes in Computer Science}, pages 47--61. Springer,
  2015.
\newblock \href {https://doi.org/10.1007/978-3-319-19458-5_4}
  {\path{doi:10.1007/978-3-319-19458-5_4}}.

\bibitem[LLN18]{DBLP:conf/isola/LarsenLN18}
Kim~G. Larsen, Florian Lorber, and Brian Nielsen.
\newblock 20 years of {UPPAAL} enabled industrial model-based validation and
  beyond.
\newblock In Tiziana Margaria and Bernhard Steffen, editors, {\em Leveraging
  Applications of Formal Methods, Verification and Validation. Industrial
  Practice - 8th International Symposium, ISoLA 2018, Limassol, Cyprus,
  November 5-9, 2018, Proceedings, Part {IV}}, volume 11247 of {\em Lecture
  Notes in Computer Science}, pages 212--229. Springer, 2018.
\newblock \href {https://doi.org/10.1007/978-3-030-03427-6_18}
  {\path{doi:10.1007/978-3-030-03427-6_18}}.

\bibitem[LLT{\etalchar{+}}19]{LLTYSG19}
Axel Legay, Anna Lukina, Louis{-}Marie Traonouez, Junxing Yang, Scott~A.
  Smolka, and Radu Grosu.
\newblock {Statistical Model Checking}.
\newblock In Bernhard Steffen and Gerhard~J. Woeginger, editors, {\em Computing
  and Software Science: State of the Art and Perspectives}, volume 10000 of
  {\em LNCS}, pages 478--504. Springer, 2019.
\newblock \href {https://doi.org/10.1007/978-3-319-91908-9_23}
  {\path{doi:10.1007/978-3-319-91908-9_23}}.

\bibitem[LP15]{LP15}
C.~Laneve and L.~Padovani.
\newblock {An algebraic theory for web service contracts}.
\newblock {\em Form. Asp. Comp.}, 27(4):613--640, 2015.
\newblock \href {https://doi.org/10.1007/s00165-015-0334-2}
  {\path{doi:10.1007/s00165-015-0334-2}}.

\bibitem[LRN{\etalchar{+}}22]{DBLP:journals/corr/abs-2203-09884}
Sascha Lehmann, Antje Rogalla, Maximilian Neidhardt, Anton Reinecke, Alexander
  Schlaefer, and Sibylle Schupp.
\newblock {Modeling $\mathbb{R}^3$ Needle Steering in {Uppaal}}.
\newblock In Clemens Dubslaff and Bas Luttik, editors, {\em Proceedings of the
  5th Workshop on Models for Formal Analysis of Real Systems (MARS@ETAPS)},
  volume 355 of {\em EPTCS}, pages 40--59, 2022.
\newblock \href {https://doi.org/10.4204/EPTCS.355.4}
  {\path{doi:10.4204/EPTCS.355.4}}.

\bibitem[LSM{\etalchar{+}}20]{DBLP:conf/nfm/LiuSMO020}
Si~Liu, Atul Sandur, Jos{\'{e}} Meseguer, Peter~Csaba {\"{O}}lveczky, and
  Qi~Wang.
\newblock Generating correct-by-construction distributed implementations from
  formal {MAUDE} designs.
\newblock In Ritchie Lee, Susmit Jha, and Anastasia Mavridou, editors, {\em
  NFM}, volume 12229 of {\em LNCS}, pages 22--40. Springer, 2020.
\newblock \href {https://doi.org/10.1007/978-3-030-55754-6_2}
  {\path{doi:10.1007/978-3-030-55754-6_2}}.

\bibitem[LSP82]{DBLP:journals/toplas/LamportSP82}
Leslie Lamport, Robert~E. Shostak, and Marshall~C. Pease.
\newblock The byzantine generals problem.
\newblock {\em {ACM} Trans. Program. Lang. Syst.}, 4(3):382--401, 1982.
\newblock \href {https://doi.org/10.1145/357172.357176}
  {\path{doi:10.1145/357172.357176}}.

\bibitem[LT89]{LT89}
N.~Lynch and M.~Tuttle.
\newblock {An Introduction to Input/Output Automata}.
\newblock {\em CWI Q.}, 2:219--246, 1989.
\newblock URL: \url{https://ir.cwi.nl/pub/18164/18164A.pdf}.

\bibitem[MNF13]{MNF13}
J.~Michaux, E.~Najm, and A.~Fantechi.
\newblock {Session types for safe Web service orchestration}.
\newblock {\em J. Log. Algebr. Program.}, 82(8):282--310, 2013.
\newblock \href {https://doi.org/10.1016/j.jlap.2013.05.004}
  {\path{doi:10.1016/j.jlap.2013.05.004}}.

\bibitem[Obj]{ObjectOutputStream}
URL:
  \url{https://docs.oracle.com/javase/7/docs/api/java/io/ObjectOutputStream.html}.

\bibitem[OPB{\etalchar{+}}21]{DBLP:conf/facs2/OrlandoPBLT21}
Simone Orlando, Vairo~Di Pasquale, Franco Barbanera, Ivan Lanese, and Emilio
  Tuosto.
\newblock {Corinne, a Tool for Choreography Automata}.
\newblock In Gwen Sala{\"{u}}n and Anton Wijs, editors, {\em FACS}, volume
  13077 of {\em LNCS}, pages 82--92. Springer, 2021.
\newblock \href {https://doi.org/10.1007/978-3-030-90636-8_5}
  {\path{doi:10.1007/978-3-030-90636-8_5}}.

\bibitem[PMT24]{DBLP:conf/coordination/PomboMT24}
Carlos Gustavo~L{\'{o}}pez Pombo, Pablo Montepagano, and Emilio Tuosto.
\newblock Search: An execution infrastructure for service-based software
  systems.
\newblock In Ilaria Castellani and Francesco Tiezzi, editors, {\em Coordination
  Models and Languages - 26th {IFIP} {WG} 6.1 International Conference,
  {COORDINATION} 2024, Held as Part of the 19th International Federated
  Conference on Distributed Computing Techniques, DisCoTec 2024, Groningen, The
  Netherlands, June 17-21, 2024, Proceedings}, volume 14676 of {\em Lecture
  Notes in Computer Science}, pages 314--330. Springer, 2024.
\newblock \href {https://doi.org/10.1007/978-3-031-62697-5_17}
  {\path{doi:10.1007/978-3-031-62697-5_17}}.

\bibitem[PPN{\etalchar{+}}23]{DBLP:journals/corr/abs-2310-20395}
Jos{\'{e}} Proen{\c{c}}a, David Pereira, Giann~Spilere Nandi, Sina Borrami, and
  Jonas Melchert.
\newblock {Spreadsheet-based Configuration of Families of Real-Time
  Specifications}.
\newblock In Maurice~H. ter Beek and Clemens Dubslaff, editors, {\em
  Proceedings of the 1st Workshop on Trends in Configurable Systems Analysis
  (TiCSA@ETAPS)}, volume 392 of {\em EPTCS}, pages 27--39, 2023.
\newblock \href {https://doi.org/10.4204/EPTCS.392.2}
  {\path{doi:10.4204/EPTCS.392.2}}.

\bibitem[RCS{\etalchar{+}}22]{Roggenbach2022}
Markus Roggenbach, Antonio Cerone, Bernd{-}Holger Schlingloff, Gerardo
  Schneider, and Siraj~Ahmed Shaikh.
\newblock {\em Formal Methods for Software Engineering: Languages, Methods,
  Application Domains}.
\newblock TTCS. Springer, 2022.
\newblock \href {https://doi.org/10.1007/978-3-030-38800-3}
  {\path{doi:10.1007/978-3-030-38800-3}}.

\bibitem[RG21]{Rai_Gangadharan_2021}
Gopal~N. Rai and G.~R. Gangadharan.
\newblock Model checking based web service verification: A systematic
  literature review.
\newblock {\em IEEE Transactions on Services Computing}, 14(3):747–764, 2021.
\newblock \href {https://doi.org/10.1109/tsc.2018.2845401}
  {\path{doi:10.1109/tsc.2018.2845401}}.

\bibitem[RW87]{RW87}
Peter~J. Ramadge and Walter~M. Wonham.
\newblock Supervisory control of a class of discrete event processes.
\newblock {\em SIAM J. Control Optim.}, 25(1):206--230, 1987.
\newblock \href {https://doi.org/10.1137/0325013} {\path{doi:10.1137/0325013}}.

\bibitem[SF17]{DBLP:journals/corr/SainiF17}
Shruti Saini and Ansgar Fehnker.
\newblock {Evaluating the Stream Control Transmission Protocol Using {Uppaal}}.
\newblock In Holger Hermanns and Peter H{\"{o}}fner, editors, {\em Proceedings
  of the 2nd Workshop on Models for Formal Analysis of Real Systems
  (MARS@ETAPS)}, volume 244 of {\em EPTCS}, pages 1--13, 2017.
\newblock \href {https://doi.org/10.4204/EPTCS.244.1}
  {\path{doi:10.4204/EPTCS.244.1}}.

\bibitem[SVW20]{DBLP:conf/sefm/Shokri-Manninen20}
Fatima Shokri{-}Manninen, J{\"{u}}ri Vain, and Marina Wald{\'{e}}n.
\newblock {Formal Verification of {COLREG}-Based Navigation of Maritime
  Autonomous Systems}.
\newblock In Frank~S. de~Boer and Antonio Cerone, editors, {\em SEFM}, volume
  12310 of {\em LNCS}, pages 41--59. Springer, 2020.
\newblock \href {https://doi.org/10.1007/978-3-030-58768-0_3}
  {\path{doi:10.1007/978-3-030-58768-0_3}}.

\bibitem[SY86]{DBLP:journals/tse/StromY86}
Robert~E. Strom and Shaula Yemini.
\newblock Typestate: {A} programming language concept for enhancing software
  reliability.
\newblock {\em IEEE Trans. Softw. Eng.}, 12(1):157--171, 1986.
\newblock \href {https://doi.org/10.1109/TSE.1986.6312929}
  {\path{doi:10.1109/TSE.1986.6312929}}.

\bibitem[tBBG07]{BBG07}
Maurice~H. ter Beek, Antonio Bucchiarone, and Stefania Gnesi.
\newblock {Web Service Composition Approaches: From Industrial Standards to
  Formal Methods}.
\newblock In {\em Proceedings of the 2nd International Conference on Internet
  and Web Applications and Services (ICIW'07)}. IEEE, 2007.
\newblock \href {https://doi.org/10.1109/ICIW.2007.71}
  {\path{doi:10.1109/ICIW.2007.71}}.

\bibitem[TMR20]{DBLP:journals/corr/abs-2009-08769}
Andr{\'{e}} Trindade, Jo{\~{a}}o Mota, and Ant{\'{o}}nio Ravara.
\newblock {Typestates to Automata and back: a tool}.
\newblock In Julien Lange, Anastasia Mavridou, Larisa Safina, and Alceste
  Scalas, editors, {\em Proceedings of the 13th Interaction and Concurrency
  Experience (ICE)}, volume 324 of {\em EPTCS}, pages 25--42, 2020.
\newblock \href {https://doi.org/10.4204/EPTCS.324.4}
  {\path{doi:10.4204/EPTCS.324.4}}.

\bibitem[WVHFM06]{WITSCH2006101}
Daniel Witsch, Birgit Vogel-Heuser, Jean-Marc Faure, and Gaëlle Marsal.
\newblock Performance analysis of industrial ethernet networks by means of
  timed model-checking.
\newblock In {\em INCOM}, volume~39, pages 101--106, 2006.
\newblock \href {https://doi.org/10.3182/20060517-3-FR-2903.00063}
  {\path{doi:10.3182/20060517-3-FR-2903.00063}}.

\bibitem[WWH{\etalchar{+}}22]{DBLP:journals/ijcomsys/WangWHLGCG22}
Jie Wang, Xintao Wu, Gang Hou, Pengfei Li, Ao~Gao, Zhichao Chen, and Haoyu Gao.
\newblock Modeling and reliability verification of industrial control network
  protocol based on time state transition matrix.
\newblock {\em Int. J. Commun. Syst.}, 35(9), 2022.
\newblock \href {https://doi.org/10.1002/dac.5140}
  {\path{doi:10.1002/dac.5140}}.

\end{thebibliography}
